\documentstyle[11pt,aaspp4]{article}

\def\Min{\hbox{${}^{\prime}$\llap{.}}}
\def\Sec{\hbox{${}^{\prime\prime}$\llap{.}}}

\def\deg{\hbox{${}^\circ$}}
\def\min{\hbox{${}^{\prime}$}}
\def\sec{\hbox{${}^{\prime\prime}$}}
\def\kms{km s$^{-1}$}

\lefthead{Ferrarese et al.}
\righthead{A New Distance to NGC 2541}

\begin{document}

The HST Key Project on the Extragalactic Distance Scale XII. 
The Discovery of Cepheids and a New Distance to NGC 2541

Laura Ferrarese, Fabio Bresolin, Robert C. Kennicutt, Jr., Abhijit Saha, 
Peter B. Stetson, Wendy L. Freedman, Jeremy R. Mould, Barry F. Madore, 
Shoko Sakai, Holland C. Ford, Brad K. Gibson, John A. Graham, 
Mingsheng Han, John G. Hoessel, John Huchra, Shaun M. Hughes, 
Garth D. Illingworth, Randy Phelps, Charles F. Prosser and N.A. Silbermann

\footnotetext{Based on observations with the NASA/ESA {\it Hubble Space
Telescope}, obtained at the Space Telescope Science Institute, which is
operated by AURA, Inc. under NASA Contract NO. NAS5-26555.}

\altaffiltext{2}{Hubble Fellow}
\altaffiltext{3}{California Institute of Technology, Pasadena CA 91125, USA}
\altaffiltext{4}{Steward Observatories, University of Arizona, Tucson AZ 85721,
USA}
\altaffiltext{5}{European Southern Observatory, Garching, Germany}
\altaffiltext{6}{Kitt Peak National Observatory, NOAO, Tucson AZ 85726, USA}
\altaffiltext{7}{Dominion Astrophysical Observatory, Victoria, British Columbia
V8X 4M6, Canada}
\altaffiltext{8}{Carnegie Observatories, Pasadena CA 91101, USA}
\altaffiltext{9}{Mount Stromlo and Siding Spring Observatories, Institute of
Advanced Studies, ANU, ACT 2611, Australia}
\altaffiltext{10}{NASA/IPAC Extragalactic Database and California Institute of
Technology, Pasadena CA 91125, USA}
\altaffiltext{11}{Johns Hopkins University and Space Telescope
Science Institute, Baltimore MD 21218, USA}
\altaffiltext{12}{Department of Terrestrial Magnetism, Carnegie Institution of
Washington, Washington DC 20015, USA}
\altaffiltext{13}{University of Wisconsin, Madison WI 53706, USA}
\altaffiltext{14}{Harvard Smithsonian Center for Astrophysics, Cambridge MA 02138
USA}
\altaffiltext{15}{Royal Greenwich Observatory, Cambridge CB3 OHA, UK}
\altaffiltext{16}{Lick Observatory, University of California, Santa Cruz CA 95064
USA}

\begin{abstract}

We report the detection of Cepheids and a new distance to the spiral
galaxy NGC 2541, based on data obtained with the Wide Field and
Planetary Camera 2 on board the Hubble Space Telescope (HST). A total
of 25 exposures (divided into 13 epochs) are obtained using the F555W
filter (transformed to Johnson V), and nine exposures (divided into
five epochs) using the F814W filter (transformed to Cousins I).
Photometric reduction of the data is performed using two independent
packages, DoPHOT  and DAOPHOT II/ALLFRAME, which give very good
agreement in the measured magnitudes. A total of 34 bona fide
Cepheids, with periods ranging from 12 to over 60 days, are identified
based on both sets of photometry.  By fitting V and I
period-luminosity relations, apparent distance moduli are derived
assuming a Large Magellanic Cloud distance modulus and mean color
excess of 18.50 +/- 0.10 mag and E(B-V) = 0.10 mag respectively.
Adopting A(V)/E(V-I)=2.45, we  obtain a true distance modulus to NGC
2541 of 30.47 +/- 0.11 (random) +/- 0.12 (systematic) mag (D = 12.4
+/- 0.6 (random) +/- 0.7 (systematic) Mpc), and a total (Galactic plus
internal) mean color excess E(B-V) = 0.08 +/- 0.05 (internal error)
mag.

\end{abstract}


\section{Introduction}

This paper presents, analyzes and discusses Hubble Space Telescope
(HST) $V$ and $I$ observations of the Sa(s)cd (RC3, de Vaucouleurs et al.
1991) galaxy NGC 2541. These were obtained over a 47 day interval for
the purpose of discovering and measuring Cepheid variables, and then
using their period$-$luminosity (PL) relation to determine a Cepheid
distance to their host system.

The observations are part of the HST Key Project on the Extragalactic
Distance Scale (Freedman et al.  1994a, 1994b, 1994c; Kennicutt,
Freedman \& Mould 1995). Cepheid distances measured for the 18 Key
Project galaxies, all within a redshift of approximately 1500 km
s$^{-1}$, will provide an accurate absolute calibration for a number of
secondary distance indicators, such as the Tully$-$Fisher relation, the
planetary nebula luminosity function, the surface brightness
fluctuation method, the expanding photosphere method for Type II supernovae, the globular
cluster luminosity function and the type Ia supernova standard candle.
The ultimate goal of the Key Project is to employ these
distance indicators to determine the Hubble constant, {\it H$_0$}, to
within 10\% accuracy.

NGC 2541, with an inclination of 58\deg~(Bottinelli et al. 1985b), was
selected as part of the Key Project because of its potential as a
Tully$-$Fisher calibrator. Given previous estimates of its distance
(e.g. $\sim 8$ Mpc, Bottinelli et al. 1986) NGC 2541 was considered to be a
relatively easy target for the detection and measurement of Cepheid
variables using HST. In addition, the galaxy has the same metallicity
as the LMC (which we use as calibrator for the zero points of the
Cepheid PL relation), therefore the derivation of its Cepheid distance
bypasses any complications associated with a dependence of the PL
relation on metal abundance (e.g. Kennicutt et al.  1998).

The galaxy (RA = 08h14m40s, Dec = +49\deg03\min44\sec~ at equinox 2000)
belongs to the NGC 2841 group (de Vaucouleurs 1975), which occupies a
15\deg $\times$ 7\deg~ region of the sky near the border of Ursa Major
and Lynx.  Besides the loose triplet formed by NGC 2541, NGC 2500 and
NGC 2552, the group is comprised of four additional large spirals and,
possibly, several dwarf systems. All of the members have low systemic
velocities, in the range 420 to 750 \kms; in particular, the 21 cm
velocity for NGC 2541 is $556 \pm 4$ \kms (RC3, de Vaucouleurs et al.
1991).

NGC 2541 is the eleventh Key Project galaxy for which a distance has been
determined; published results include M81 (Freedman et al. 1994a), M100
(Freedman et al.  1994c, Ferrarese et al. 1996), M101 (Kelson et al.
1996, 1997), NGC 925 (Silbermann et al. 1996), NGC 3351 (Graham et al.
1997), NGC 3621 (Rawson et al. 1997), NGC 7331 (Hughes et al. 1998),
NGC 2090 (Phelps et al. 1998), NGC 1365 (Silbermann et al. 1998) and
NGC 4414 (Turner et al. 1998).

These measurements further expand the existing database of Cepheid
distances, measured by a number of collaborations using both HST (NGC
4603, Zepf et al. 1997; M96, Tanvir et al. 1995; NGC 4639, Saha et al.
1997; NGC 4536, Saha et al. 1996a;  NGC 4496A, Saha et al. 1996b; IC
4182, Saha et al.  1994; NGC 5253, Saha et al. 1995) and ground based
facilities (SMC and LMC, Welch et al. 1987; M31, Freedman and Madore
1990; M33, Freedman et al.  1991; IC 1613, Freedman 1988; NGC 2366,
Tolstoy et al. 1995; GR 8, Tolstoy et al. 1995; NGC 6822,
McAlary et al. 1983; IC10, Saha et al. 1996; WLM, Madore \& Freedman
1991; NGC3109, Capaccioli et al.  1992; as well as for more distant
galaxies, including NGC 2403 and the M81 group, Freedman \& Madore
1988; NGC 300, Freedman et al. 1992; Sextans A, Madore \& Freedman
1991; Sextans B, Sakai, Madore \& Freedman 1997; Leo I, Hodge \& Wright
1978; Leo A, Hoessel et al.  1994; Pegasus, Hoessel et al. 1990; M101,
Alves \& Cook 1995; and the Virgo galaxy NGC 4571, Pierce et al.
1994).

This paper is organized as follows:  \S 2 describes the data and the
preliminary reduction. \S 3 and \S 4 deal with the photometric
reduction and the variable star search. The $V$ and $I$ PL relations
and the apparent distance moduli are discussed in \S 5, while \S 6
deals with the extinction and the true distance modulus.  Discussion
and conclusions can be found in \S 7.

\section{Observations and Data Reduction}

NGC 2541 was first observed as part of the HST Key Project on 1994 December 
28, when the Wide Field and Planetary Camera 2 (WFPC2, Biretta et
al. 1994) was used to obtain two F555W (close to Johnson $V$) and two
F814W (close to Cousin $I$) images, with the purpose of surveying the
field and assessing the correctness of the exposure times. A 47 day
long observing sequence, specifically designed for finding Cepheid
variables, began almost one year later, on 1995 October 30, and
comprised a total of 25 F555W images, divided among 12 epochs, and
nine F814W images, divided among four epochs\footnotemark. The time
interval between subsequent epochs was chosen as to maximize the
probability of detecting Cepheids with periods between 10 and 60 days,
allowing at the same time for an optimum sampling of the light curves
and reducing the possibility of aliasing (Freedman et al. 1994a).  The
observation log is given in Table 1.  The observations were dithered by
an integer number of pixels (up to four) to minimize the effects of
dead or hot pixels, while the roll angle of the telescope was
maintained the same throughout the sequence. The telescope was always
guiding in fine lock, which gives a nominal pointing stability of about
3 mas.

\footnotetext{Multiple exposures were obtained at each epoch to
facilitate identification and removal of cosmic rays.}

Figure 1 shows the WFPC2 field of view superimposed on a $V$ band image
of NGC 2541 obtained at the Fred L. Whipple Observatory (FLWO) 1.2$-$m
telescope.  Each of the three Wide Field Camera (WFC) chips has a scale
of 0.10 arcsec/pixel and a field of view of $\sim 1\Min25\times1\Min25$,
while the Planetary Camera (PC) chip has a scale of 0.046 arcsec/pixel
and a field of view of $\sim 33\sec\times31\sec$.  The gain and readout noise
for each chip are about 7 e$^-$/DN and 7 e$^-$ respectively.

HST/WFPC2 data are routinely calibrated using a standard pipeline
maintained by the Space Telescope Science Institute (STScI).  The
reduction steps (described in detail by Holtzman et al. 1995a) are
performed in the following order: correction of small A/D errors;
subtraction of a bias level for each chip; subtraction of a superbias
frame; subtraction of a dark frame; correction for shutter shading
effects and division by a flat field.  We performed a few additional
non$-$standard processing on our images: the vignetted edges of the
CCD are blocked out; bad pixels and columns are also masked using the
data quality files provided by the standard pipeline, and the images
are multiplied by a pixel area map to correct for the WFPC2 geometric
distortion. Finally, all images are multiplied by four and converted to
integer format.

As for all other Key Project galaxies, `long exposure' zero points are
used to calibrate the photometry (Hill et al. 1998, see also \S 3). While this removes
the mean effect of the charge transfer efficiency (CTE) problem
(Whitmore \& Heyer 1997, Whitmore 1998), the pixel$-$to$-$pixel
dependence of the CTE is not corrected. For the NGC 2541 observations,
we estimate CTE losses to range from zero to 1.5\% (0.016 mag) going
from the bottom to the top of each chip. This effect, even if
systematic, is negligible when compared to other sources of errors in
the photometry (see \S 6.1). Note that the NGC 2541
observations were taken before the severe increase of the CTE effect 
which has been registered in the past two years (Whitmore 1998).

\section{Photometric Reduction}

Photometric analysis of the data was performed independently using
DAOPHOT II and ALLFRAME (Stetson 1994), and a variant of DoPHOT
especially formulated to deal with the peculiarities of HST data and
PSFs (Schechter  et al. 1993, Saha et al. 1994).  As extensively
discussed in Hill et al. (1998) and Ferrarese et al. (1998), DoPHOT and
DAOPHOT II/ALLFRAME use radically different approaches to solve the
complicated problem of measuring magnitudes of stellar objects in
crowded fields. Thus comparing the DoPHOT and ALLFRAME outputs
provides a powerful tool for revealing systematic errors in the
measured magnitudes that could easily go unnoticed if only one of the
two programs was used.

Both DoPHOT and ALLFRAME rely on a master star list to identify objects
in each frame. The list is created from a very deep image obtained by
combining all of the frames, and rejecting cosmic rays in the process, 
as described later.  The master star list contains our best
approximation to a complete list of stars expected in each frame. At
this point, DoPHOT and ALLFRAME follow very different routes: while
DoPHOT is run independently on each image, ALLFRAME simultaneously
reduces all images, identifying as real only objects belonging to the
master star list and appearing in a significant fraction of the frames.
This makes ALLFRAME very robust in identifying cosmic ray (CR) events,
while DoPHOT is more easily run on images from which CR have been
already removed.  Flagging CR hits is easily done since two images,
taken one immediately after the other with the same (within 0.1 pixels)
pointing, are available for each epoch.  The two images are combined
and cosmic rays are flagged by comparing the difference in values between
pairs of corresponding pixels (after accounting for possible
differences in the sky level due to changes in orbital position of the
spacecraft between subsequent exposures) to a local sigma calculated
from the combined effects of Poisson statistics and local noise. Pixels
differing by more than four sigma are flagged as cosmic rays and
replaced in the combined image by the lowest of the two input values.
Because of the severe under$-$sampling of the PC and, in particular, of
the WFC PSF, particular care is taken in assuring that tips of bright
stars are not erroneously identified as CR hits. In summary, DoPHOT is
run on a total of 13 F555W and five F814W combined images, while
ALLFRAME is run on the original 27 F555W and 11 F814W images.

Both DoPHOT and ALLFRAME compute `PSF magnitudes' for each identified
star.  The DoPHOT PSF magnitudes are proportional to the height of the
fitted PSFs. These differ from the integrated magnitudes (flux
subtended by the PSF) by an amount (the `aperture correction')
which is a constant for a given chip of a given epoch, but will change from
chip to chip and epoch to epoch due to PSF variations depending on telescope focus,
jitter, etc. DoPHOT aperture corrections are calculated as the mean
difference between aperture magnitudes (integrated within 0\Sec22 for
the PC and 0\Sec5 for the WFC) and PSF magnitudes for all bright,
isolated stars with errors on the aperture magnitudes less than 0.05
mag and converging growth curves (i.e.  aperture magnitudes must change
by no more than 0.05 mag for one pixel increments between 5 and 8
pixels). In the un$-$crowded NGC 2541 field, the number of
stars meeting these criteria varies between a few (F555W/WF4) and about
80 (F814W/WF2). The DoPHOT aperture corrections calculated for the
1995 October 30 epoch are listed in Table 2; to insure a uniform photometric
system, aperture corrections for each other epoch are calculated as the
mean difference between the aperture corrected magnitudes for the
1995 October 30 epoch and the PSF magnitudes for the epoch in question.

Finally, the DoPHOT magnitudes obtained by summing the aperture
corrections (AC) to the PSF magnitudes ($m_{PSF}$) can be converted
first to the Holtzman et al.  (1995b) `ground system' F555W and F814W
magnitudes ($m_g$) and then to $V$ and $I$ magnitudes ({\sf M}) using the
zero points (ZP, normalized to a unit exposure time) and color
corrections listed in Table 2 (see Hill et al.  1998 and Holtzman et
al.  1995b for additional details), and scaling by the exposure time
$t$ appropriate for each frame:

$${\sf M} = m_{PSF} + AC + ZP + 2.5{\rm log}t + C2(V-I) + C3(V-I)^2 = m_g + C2(V-I) + C3(V-I)^2. \eqno(1)$$

The zero points $ZP$, listed in Table 2, include the long exposure
corrections (0.05 mag in both $V$  and $I$) described by Hill et al.
(1998), and a $2.5\times{\rm log}(4.0)$ correction since the images
have been multiplied by four before being converted to integer format.

ALLFRAME aperture corrections were derived from 10$-$20 bright,
isolated stars for each chip, each filter {\it and} each frame, as the
mean difference between the ALLFRAME PSF magnitudes and the 0\Sec5
aperture magnitudes. The latter were determined with a growth-curve
analysis obtained with the program DAOGROW (Stetson 1990).  Equation
(1), with the coefficients listed in Table 2, also provides photometric
calibration for the ALLFRAME magnitudes (see Hill et al. 1998 and
Silbermann et al. 1997 for additional details). The ALLFRAME aperture
corrections for the first and second exposure of the 1995 October 30 epoch
are listed in Table 2; because of focus and jitter changes, aperture
corrections for other epochs can differ by a few hundredth of a magnitude.

\subsection{Comparison of the DoPHOT and ALLFRAME Photometry}

The comparison between the DoPHOT and ALLFRAME magnitudes is shown in
Figure 2 for a set of bright isolated secondary standard stars in each chip. 
A subset of these stars was used to derive the DoPHOT and ALLFRAME 
F555W aperture corrections as described in \S3, while additional 
stars were used for the F814W aperture corrections. The positions
and DoPHOT photometry of the secondary standard stars are tabulated in Tables A6a$-$d of
the Appendix. Figure 2 plots the difference in magnitudes between the
two sets of photometry as a function of the DoPHOT magnitudes, for
F555W and F814W respectively (note that, since the color correction
coefficients from Table 2 are the same for ALLFRAME and DoPHOT,
differences in F555W magnitudes, or F814W magnitudes, correspond to
differences in $V$, or $I$, magnitudes). The weighted means and errors
of the DoPHOT$-$ALLFRAME offsets for the secondary standard stars are
tabulated in Table 3. The agreement between the two sets of photometry
is excellent and well within the uncertainties associated with the
aperture corrections (Table 2) and with the photometric errors,
which are of the order of 0.03 mag for a 24th F555W magnitude star. 
The mean DoPHOT$-$ALLFRAME offsets and
errors derived for the Cepheid stars are tabulated in Table 4 and
plotted in Figure 3, and will be discussed in the next section.

\section{Variable Star Search}

\subsection{DoPHOT Selection Criteria for Variability}

The search for variable stars was performed on the $V$ band images,
following the procedure described in Saha \& Hoessel (1990), the main
points of which are summarized below.  We required that a star be
detected in at least 10 of the 13 F555W frames to be checked for variability.
We also excluded all stars in crowded regions by rejecting candidates
with a companion contributing more than 50\% of the total light within
a two pixel radius. Each star meeting these requirements was first
tested for variability using a $\chi^2$ test. The reduced $\chi^2_r$ is
defined by

$$\chi^2_r = {1 \over (n-1)} \sum_i^n{(m_i-\bar{m})^2 \over \sigma_i^2}, \eqno(2)$$

\noindent where $m_i$ and $\sigma_i$ are the magnitude and rms error of
a particular star as measured in the $i$-th epoch, $\bar{m}$ is the
magnitude of the star averaged over all epochs, and $n$ is the number
of epochs in which the star is detected. A star was always flagged as
variable if $\chi^2_r \ge 8$.  Stars shown as variables at a 99\%
confidence level (as defined in Saha \& Hoessel) but with
$\chi^2_r < 8$ were checked for periodicity using a variant of the
Lafler \& Kinman (1965) method of phase dispersion minimization, in
the period range between 3 and 100 days.  Stars with $\Lambda \ge 3$
were flagged as variables, where $\Lambda$ is as used in Saha \&
Hoessel (1990), following the definition by Lafler \& Kinman.

Several spurious variables are detected in this procedure, as a
consequence of non-Gaussian sources of error (Saha \& Hoessel 1990),
various anomalies in the images (e.g. residual cosmic ray events), and
crowding. Therefore each star selected on the basis of the $\chi^2_r$
and the $\Lambda$ tests was visually inspected by blinking several of
the individual frames against each other. This allowed us to select
from the original list of 289 candidates a total of 56 variable stars. The best period
for each variable was selected by phasing the data for all periods
between 3 and 100 days in incremental steps of 0.01 days.  Although in
most cases the final period adopted corresponds to the minimum value of
the phase dispersion, in a few cases an obvious improvement of the
light curve was obtained for a slightly different period.  Because of
the careful sampling of the data, aliasing does not present a serious
problem, and we were always able to determine a preferred period for
all variables, except for those with period exceeding the length of the
observing window.  For these long period variables, the one$-$year
pre$-$visit helps in narrowing the possible ranges of acceptable periods,
but does not remove the degeneracy completely.

\subsection{ALLFRAME Selection Criteria for Variability}

A total of 45 variables are identified based on the ALLFRAME
photometry, using two independent methods. The first method searches
for stars with unusually high dispersion in the mean $V$ magnitudes,
and is equivalent to the method described in \S 4.1.  The second method
employs a variation of the correlated variability test by Welch \&
Stetson (1993). Suitable candidates are defined by having the ratio
$\sigma$ of the average absolute deviation from the mean to the mean
error in excess of 1.3.  Periods for the candidate variables are found
using a phase$-$dispersion minimization routine as described by
Stellingwerf (1978). The resulting light curves are checked by eye to
verify the best period for each candidate.

\subsection{Definition of the Final Cepheids Sample}

To be included in our final list of `bona fide' Cepheids, we require a
star to be detected and meet the variability criteria for both ALLFRAME
and DoPHOT.  Cepheids have highly periodic and very distinctive light
curves. In the period range we are interested in, they are
characterized by a steep rise and slower (by over a factor two for the
shorter period Cepheids) decline, and a narrow maximum and broader
minimum. These asymmetries are less pronounced the longer the period (a
relationship known as the Hertzsprung progression), but only the
longest period Cepheids (100 days or more) have nearly symmetric light
curves. The magnitude range spanned is of the order of one magnitude in
$V$, increasing slowly with period. For these reasons, variables with
sinusoidal light curves or small amplitude are to be regarded
suspiciously and are not included in our final Cepheid sample.
Also, as already mentioned in \S 4.1, variables found in very crowded
regions (i.e. contributing to less than 50\% of the sky$-$subtracted light within
a 2 pixel radius) are rejected.

Adopting these criteria yields 44 variables in common between the
DoPHOT and ALLFRAME lists; of these, 34 show convincing Cepheid$-$like
light curves based on both sets of photometry. These stars are listed
in Table 5, numbered in order of decreasing period. The
epoch$-$by$-$epoch DoPHOT photometry for each Cepheid is given in Table
6 for F555W and Table 7 for F814W; we do not tabulate the ALLFRAME
photometry for any of the variable stars, this is however available on
line through the HST Key Project Archives at
http://www.ipac.caltech.edu/H0kp.  A list of variables that were not
included in the final Cepheid list, along with detailed reasons for the
exclusion, is given in Table A1 of the Appendix.

The 34 newly discovered Cepheids and the additional variables are
identified in each of the WFPC2 chips in Figures 4a-d.  Finding charts
for the Cepheids are given in Figures 5. These finding charts cover a
$5\sec \times 5\sec$ region and have the same orientation as the
corresponding chips displayed in Figures 4a-d.  The contrast and
intensity level have been adjusted differently for each finding chart,
therefore the relative brightness of the Cepheids cannot be inferred
from them. The DoPHOT light curves for all Cepheids are shown in
Figures 6.  Finding charts and light curves for the remaining NGC 2541
variables are shown in the Appendix.

\subsection{Mean Magnitudes}

In line with all other papers in this series (e.g. Ferrarese et al.
1996), we calculate both intensity averaged $m_{int}$ and phase
weighted $m_{ph}$ mean magnitudes for all Cepheids:

$$m_{int} = -2.5 \log_{10}\sum_{i=1}^n {1 \over n} 10^{-0.4\times m_i}, \eqno(3)$$

$$m_{ph} = -2.5\log_{10}\sum_{i=1}^n 0.5(\phi_{i+1} - \phi_{i-1})10^{-0.4
\times m_i}, \eqno(4)$$

\noindent where $n$ is the total number of observations, and $m_i$ and
$\phi_i$ are the magnitude and phase of the $i$-th observation in order
of increasing phase.

Differences between $m_{int}$ and $m_{ph}$ are more pronounced when
the phase coverage is less uniform. Typical absolute
differences are of the order 0.01$-$0.02 magnitudes, but reach 0.1
magnitudes in a few cases of not very well sampled light curves (e.g.
C4 and C20).  The mean difference between DoPHOT phase weighted and
intensity averaged magnitudes is $-0.01 \pm 0.05$ for F555W and $0.00
\pm 0.04$ for F814W.

Because of the poor phase coverage of the F814W light curves, an
additional correction to both intensity averaged and phase weighted
magnitudes is needed, as described in Freedman et al. (1994a).  Following
the finding that a Cepheid $V$ light curve can be mapped into its $I$
light curve by simple scaling by a factor $\alpha = 0.51$ (Freedman
1988), this correction is equal to $\alpha$ times the difference
between the mean F555W magnitude obtained using the complete (up to 13
points) data set and the F555W magnitude calculated using only the data
points in common with the F814W observations.  The
result (which needs to be calculated separately for intensity averaged
and phase weighted magnitudes) is then summed to the mean F814W
magnitudes.  Typical absolute values are 0.01$-$0.02
magnitudes, but they range between -0.1 to 0.1 magnitudes, being more
substantial for the less uniformly sampled F814W light curves.

The final list of Cepheids, their periods, F555W, F814W (corrected as
described above), $V$ and $I$ intensity averaged and phase weighted
mean magnitudes are listed in Table 8 for DoPHOT and Table 9 for
ALLFRAME. The corrections adopted for the F814W magnitudes, calculated
as described above, are shown in parenthesis, also the reduced
$\chi^2_r$ for DoPHOT (\S 4.1) and $\sigma$ for ALLFRAME (\S 4.2) are
listed in the last column of the tables.

The agreement in the periods determined independently from the DoPHOT
and ALLFRAME photometry for the 34 Cepheids is excellent (see Tables 8
and 9), often within 2$-$3\% or better. This is shown in Figure 7.
However, note that in spite of the fact that all DoPHOT$-$ALLFRAME period
differences are contained within $\pm$ 10\%, this is only a lower limit
to the real uncertainty in the periods, since it does not quantify
the possibility of aliasing, in particular for the longer period
Cepheids.\footnotemark

\footnotetext{At periods longer than $\sim 45$ days, the Lafler$-$Kinman
plot is characterized by a wide trough with no wall at the long period
end. The effect of the one year pre$-$visit is to resolve the trough in
a series of narrow minima, but the degeneracy is not removed.}

The DoPHOT and ALLFRAME phase weighted mean magnitudes for the common
Cepheids are compared in Figure 3, and the mean differences are
tabulated in Table 4. Within the quoted errors, the WF chips are
consistent with zero offset. The Cepheids in the PC field differ by
five hundredth of a magnitude, or two to three times the quoted
uncertainty in the mean. However, this difference is not significant when
compared to errors in the photometry and aperture corrections (see \S 3
and \S 6.1).

Before proceeding to use our Cepheid sample to derive apparent and true
distance moduli to NGC 2541, it is wise to perform two further checks.
The first is to make sure that the photometric errors for the Cepheids
are similar, in the mean, to those of non$-$variable stars with
comparable magnitude.  One concern addressed with this point (to be
discussed in more detail in Ferrarese et al. 1998, which will deal with
the photometric recovery of artificial stars in crowded fields), is
that some of the Cepheids might be part of (not necessarily physical)
pairs which are not resolved by DoPHOT and/or ALLFRAME, so that their
magnitudes are measured too bright. Photometric errors in this case
would be higher than expected for an isolated star of similar
magnitude, since the contamination by a companion is likely to give a poorer PSF fit.  The DoPHOT photometric errors are shown in Figure 8 for the
1995 October 30 epoch and WF2; these are representative of the errors for all
other epochs, and since WF2 is the most crowded chip, they are upper
limits to the errors measured in the other chips.  The Cepheids,
plotted as solid dots surrounded by large circles, lie on the main
error ridge$-$line, with the exception of two outliers: C2 which is in
a crowded region in WF3 (therefore a larger error is to be expected),
and C8 which was measured with a large uncertainty in the 1995 October 30
epoch, but normal (for its magnitude) uncertainty in all other epochs.
The second test is to confirm that all variables listed as Cepheids lie
within the instability strip in a color magnitude diagram (CMD). An
$I$, $V-I$ CMD is shown in Figure 9; Cepheids are marked by the large
circles, while all other stars are plotted as points. The
Cepheids do in fact lie in a band between $0.4 \leq V-I \leq 1.4$,
fully within the instability strip.

\section{The Period-Luminosity Relation and the Apparent Distance Moduli 
to NGC 2541}

We will first consider the apparent distance moduli derived from the
DoPHOT photometry.  In  keeping with all other papers in this series,
the $V$ and $I$ PL relations derived by Madore \& Freedman (1991) for
a sample of 32 Cepheids in the LMC, provide the zero points for
calibration of the apparent $V$ and $I$ distance moduli to NGC 2541.
Assuming an LMC true modulus and average line-of-sight reddening  of
$18.50 \pm 0.10$ and  $E(B-V) = 0.10$ mag respectively, these PL
relations give for the absolute magnitude $M$ of Cepheids as a function
of period $P$:

$$M_V  = -2.76[\log_{10}P - 1.0] - 4.16 \eqno(5)$$ 

\noindent and 

$$M_I = -3.06[\log_{10}P - 1.0] - 4.87. \eqno(6)$$

As in all other papers in this series, in fitting the NGC 2541 data,
the slope of the PL relation (both in $V$ and $I$) is fixed to the LMC
values given in equations (5) and (6). This minimizes those biases due
to incompleteness at short periods in the NGC 2541 sample, which would
artificially produce a shallower PL slope.  Because Cepheids C1 through
C6 (Table 8) do not have well defined periods, they are not used in
fitting the PL relation. Since our sample does not contain Cepheids
with period less than $\sim 8$ days (the shortest period Cepheid,
C34, has $P = 12$ days) we are confident that all of the Cepheids are
oscillating in the fundamental mode, rather than in the first harmonic
(Smith et al. 1992). In conclusion, the Cepheids used to fit the PL
relations have periods ranging between 12 and 47 days.

The fits to the NGC 2541 $V$ and $I$ PL relations are performed as
described in Freedman et al. 1994a. Outliers deviating from the mean 
more than three times the $\sigma$ of the best fit, in either 
the $V$ or $I$ PL plots, 
are excluded from the fit of both the $V$ and $I$ PL relations. 
The best fitting PL relations for the remaining
27 Cepheids, using phase weighted mean magnitudes are:

$$m_V  = -2.76[\log_{10}P - 1.0] + (26.58 \pm 0.04) \eqno(7)$$ 

\noindent and 

$$m_I = -3.06[\log_{10}P - 1.0] + (25.76 \pm 0.05), \eqno(8)$$

\noindent where the uncertainties are equal to the rms of the fit 
divided by the square root of the number of
Cepheids constraining the fit. 
The best fitting PL relations are
shown by the solid lines in Figure 10. The dashed lines, drawn at
$\pm0.54$ mag for the $V$ PL plot, and $\pm0.36$ mag for the $I$ PL
plot, represent the 2$\sigma$ scatter of the best fitting PL relation for 
the LMC Cepheids 
(Madore \& Freedman 1991). As for other galaxies in this series 
(eg. NGC 3621, NGC 925, M100), the $V$ and $I$ PL relations exhibit 
similar scatter. Due to the small sample size of both the LMC and 
NGC 2541 Cepheids, and to the limited phase 
coverage of the $I-$band data, this result is
not significantly discrepant with the 3:2 ratio of the $V$ and $I$ scatter 
observed by Madore \& Freedman for the LMC Cepheids.

The only 3$\sigma$ outlier, C11, is marked 
by the open
circle. Of all the Cepheids, C11 is the one with the least well defined
light curve: this appears fairly noisy and symmetric in DoPHOT 
(but much less so
in ALLFRAME). The Lafler$-$Kinman plot shows that a shorter period, about 7 
days, cannot be excluded. A posteriori, it is possible that C11 is a short 
period Cepheids with an alias at $\sim 30$ days.

The values given in equations (7) and (8), combined with the LMC
calibrating PL relations (equations 5 and 6), give apparent distance
moduli to NGC 2541 $\mu_V  = 30.74 \pm 0.04$ mag and $\mu_I = 30.63 \pm
0.05$ mag, where the quoted uncertainties only reflect the rms
dispersions in the PL fit. A complete estimate of the errors will be
given in \S 6.1.  As discussed in \S 4.4, the fact that all of the
Cepheids have very well sampled light curves gives rise to only very
small differences between phase weighted and intensity averaged
magnitudes. As a consequence, the apparent distance moduli given above
are reproduced (to within 0.01 mag in $V$ and 0.001 mag in $I$) if
intensity averaged magnitudes are used. 

A more serious source of concern is the fact that the faint end
incompleteness intrinsic to the photometry and the finite width of the
PL relation, will in general conspire to produce an {\it overall} shift
of the PL relation towards brighter magnitudes (since the slope of the
PL relation is kept fixed). Artificial stars simulations of the NGC 2541
field, to be described elsewhere (Ferrarese et al. 1998), show that
incompleteness effects are noticeable for $V > 26.5$ (about 50\% of $V
= 26.5$ mag stars are recovered). From Table 8, $V=26.5$ mag
corresponds to Cepheids with period of about 15$-$18 days. To be
precise, because our sample of Cepheids includes only objects with high
quality (small errors) photometry, a star needs, by definition, to be
in a fairly un$-$crowded field to be selected as a Cepheid, therefore the
Cepheid sample is {\it by construction} not as affected by faint end
incompleteness biases (which are mainly produced by crowding) as the
complete non$-$variable star sample, and we expect significantly more
than 50\% of $V = 26.5$ mag Cepheids to be detected. However, to be
conservative, we fit PL relations to the subsample of 15 Cepheids with
periods larger than 20 days (C7 to C21). The resulting distance moduli
are 0.02 magnitudes larger, in both $V$ and $I$, than the ones quoted
in the previous paragraph; this increase is not significant when
compared to the formal internal errors in the fits (0.04 mag in $V$
and 0.05 mag in $I$ when the smaller sample is used).  Therefore the
concerns arisen by Ferrarese et al. (1996) for the M100 Cepheids (for
which a 0.1 increase in the apparent distance moduli was observed when
the period cutoff was moved from eight to 20 days) do not apply to the
closer and less crowded NGC 2541 field.

Using the phase weighted magnitudes derived from the ALLFRAME
photometry (Table 9) gives distance moduli $\mu_V  = 30.70 \pm 0.04$
and $\mu_I = 30.62 \pm 0.04$, in agreement with the DoPHOT distance
moduli within the formal uncertainty of the fits. The final Cepheid
distance to NGC 2541 reported in this paper is derived from the DoPHOT
apparent distance moduli.

\clearpage

\section{The Extinction and the True Distance Modulus}

The true distance modulus to NGC 2541, $\mu$, is given by 

$$\mu = \mu_V - A(V) = \mu_I - A(I)\eqno(9)$$

\noindent where the $V$ and $I$ band absorption coefficients $A(V)$ and
$A(I)$ obey the relation $A(V)/E(V-I) = 2.45$. This is consistent with
the extinction laws by Cardelli, Clayton \& Mathis  (1989), Stanek
(1996) and Dean, Warren \& Cousins (1978), and assumes $R_V =
A(V)/E(B-V) = 3.3$.

Given the apparent distance moduli determined in the previous section,
we obtain from equation (9) $A(V) = 0.27 \pm 0.17$ (internal error) and
therefore $E(V-I) = 0.11 \pm 0.07$ and $E(B-V) = 0.08 \pm 0.05$
(internal errors). The true distance modulus is therefore\footnotemark

\footnotetext{As discussed in Ferrarese et al. 1996, this result is
correct if $R_V$ is the same for the LMC and NGC 2541. If this is not
the case, the distance modulus to NGC 2541 will depend on the value of
$A(V)$ for the LMC. This source of uncertainty is discussed in \S 6.1.}

$$\mu = 30.47 \pm 0.11 {\rm(random)} \pm 0.12 {\rm(systematic)}~ {\rm mag}, \eqno(10)$$

\noindent corresponding to a linear distance

$$D = 12.4 \pm 0.6 {\rm(random)} \pm 0.7 {\rm(systematic)}~ {\rm Mpc}, \eqno(11)$$

\noindent where the quoted errors will be discussed in \S 6.1.  Using
the distance moduli derived from the ALLFRAME photometry gives a true
distance modulus to NGC 2541 of 30.50 mag, and a color excess $E(B-V) =
0.06$.

\subsection{Error Budget}

The error budget in the determination of the true distance modulus is
given in Table 10. There are five main sources of uncertainty: \\
\noindent {\it i)} errors in the calibrating LMC PL relations given in
equations (5) and (6) (including errors in the LMC true distance
modulus and in the zero points of the LMC $V$ and $I$ PL relations).
Because this uncertainty affects all Key Project galaxies exactly the
same way, it is of systematic nature.\\ \noindent {\it ii)} errors in
the adopted value of $R_V$ for NGC 2541. This and all subsequent errors
propagate randomly to the entire sample of Key Project galaxies.\\
\noindent {\it iii)} Errors intrinsic to the assumption that the LMC
and NGC 2541 share the same metallicity; \\
\noindent {\it iv)} errors associated with the assumption that the
intrinsic ratio of total$-$to$-$selective absorption $R_V$ is the same
for NGC 2541 and the LMC;\\
\noindent  {\it v)} errors in the photometric calibration of the
NGC 2541 data (photometric zero points and aperture corrections, see
Table 2); and finally\\
\noindent {\it vi)} errors associated with fitting the PL relation to
NGC 2541 (equations 7 and 8).\\

A detailed discussion on the errors associated with the LMC calibrating
PL relations (point {\it i}) is given in Phelps et al. (1998) and
summarized in Table 10, and amounts to a 0.12 mag.

Varying $R_V$ within an acceptable range (e.g. Cardelli, Clayton \&
Mathis 1989) produces negligible effects on the final distance modulus.
We assumed $R_V = 3.3$ for both the LMC and NGC 2541 throughout our
derivation of the true distance modulus; if $R_V$ = 3.0 instead, then
we obtain from equation (9) $A_V = 0.26$, and the distance modulus is
underestimated by only 0.01 magnitudes.

Point {\it iii)} addresses the concern that metallicity differences
between the calibrating LMC sample and the project galaxy (NGC 2541 in
our case) can affect the absolute slope and zero point of the PL
relation (Kochanek 1998). This is not an issue for NGC 2541, which has
the same  oxygen abundances as the LMC ([O/H] = $-0.40$ for the LMC,
Kennicutt et al. 1998; [O/H] = $-0.42 \pm 0.09$ for NGC 2541, Zaritsky,
Kennicutt \& Huchra 1994). Note that the recent work by Kennicutt et
al.  (1998) found only a weak dependence of the inferred distance
modulus $\mu$ on metal abundance, hence even for metal abundances
differing at the one sigma level, the impact on the distance modulus
would be very contained. Therefore, neglecting metallicity
effects leads to overestimate the NGC 2541 distance by a mere $0.005
\pm 0.02$ mag.

Point {\it iv)} is more subtle. If the ratio of total$-$to$-$selective
absorption $R_V$ is different between the LMC and NGC 2541, the true
distance modulus to NGC 2541 depends explicitly on the value of the
visual absorption $A_V$ to the LMC (Ferrarese et al.  1996).
From equation (8) of Ferrarese et al.,
assuming $R_V = 3.0$ for NGC 2541, $R_V = 3.3$ and $E(B-V) = 0.10$ for
the LMC (foreground plus internal reddening),
leads to a true distance modulus 0.014 mag smaller than quoted in
equation (10).

Errors on the photometric calibration include errors in the aperture
corrections (Table 2, we conservatively adopt the largest of the
measured uncertainties, 0.011 mag in $V$ and 0.014 mag in $I$), in the
photometric zero points ($\pm 0.02$ mag, the long exposure zero points
are adopted, Hill et al. 1998) and in the long$-$vs$-$short exposure
correction ($\pm$ 0.02 mag, Hill et al 1998). These errors, being
uncorrelated, are added in quadrature to give a 0.03 mag random error
in both $V$ and $I$. When propagated to the true distance modulus, the
errors $\epsilon(V)$ and $\epsilon(I)$ on the $V$ and $I$ magnitudes
thus derived, are to be combined in quadrature (since they are
uncorrelated), weighted by  the  term $R=A(V)/E(V-I)$, so that the
corresponding error on the distance modulus is equal to
$\sqrt{\epsilon(V)^2\times(1-R)^2 + \epsilon(I)^2\times R^2}$ (see
equation 7 of Ferrarese et al. 1996 for further details).

Finally, errors in the $V$ and $I$ PL relations are equal to the rms
scatter of the data points with respect to the fit, divided by the
square root of  the number of Cepheids contributing to the fit. Much of
the {\it scatter} in the individual PL relations is highly correlated
due to intrinsic color and differential reddening differences from star
to star.  Further discussion on this point can be found in Phelps et
al. (1998).

\section{Discussion and Conclusions}

A few distance estimates to NGC 2541 are available in the literature and are summarized in Table 11.

Based on the RC2 $B_0^T$ magnitude and 21 cm data from Bottinelli et
al.  (1992),  Bottinelli et al.  (1984, 1985b, 1986) calculated a
distance modulus to NGC 2541 via the B$-$band Tully Fisher relation
(calibrated by Bottinelli et al. 1983).  The derived distance modulus
is 29.41 $\pm$ 0.18 mag (Bottinelli et al. 1984). Based on the same
data, but slightly different calibration, this was later revised to
29.53 $\pm$ 0.26 mag (Bottinelli et al. 1985b) and 29.82 $\pm$ 0.19 mag
(Bottinelli et al.  1986). This distance modulus disagrees significantly with the Cepheid distance derived
in this paper. However, we notice that the distance moduli given by
Bottinelli et al. (1986) for NGC 2500 and NGC 2841, likely members of
the same group, are 30.44 $\pm$ 0.91 mag and 30.54 $\pm$ 0.17 mag
respectively, close to our value for NGC 2541.  NGC 2541 forms a
triplet with NGC 2500 and NGC 2552. Their heliocentric velocities are
556 \kms, 516 \kms~and 519 \kms~ respectively (RC3), and the galaxies
are within a few degrees of each other, which strongly suggests
physical association. It seems unlikely that NGC 2541 is a foreground
object as the Bottinelli et al.  distance would imply. Note that NGC
2841 itself is only at a heliocentric velocity of 631 \kms, although it
is $\sim 11$ degrees away.

An independent IR Tully$-$Fisher distance to NGC 2541 was kindly
provided by Stephane Courteau (private communication). Based on the
Willick et al. (1997) Tully$-$Fisher calibration, the distance modulus
to NGC 2541 is 30.86 $\pm$ 0.3 mag.  Using the inverse Tully$-$Fisher
relation instead (Willick et al. 1997), a consistent result is
obtained:  30.78 $\pm$ 0.3 mag.  Within the quoted errors, these values
are in agreement with the Cepheid distance modulus found in this paper,
and strengthen the suspicion that the data used by Bottinelli et al.
(1984, 1985b, 1986) are in error.  Unfortunately, there are no
early$-$type galaxies in the vicinity of NGC 2541 that can be used to
derive an independent distance to the group using the surface
brightness fluctuation or planetary nebulae luminosity function
methods. We note that the distance moduli derived by Bottinelli et al.
(1985a) using the method of `sosie' galaxies, and by de Vaucouleurs
(1975), based on a variety of secondary distance indicators (HII
regions, brightest stars, luminosity class), are likely to suffer from
much larger uncertainties than quoted in Table 11.

The following is a brief summary of the results presented in this
paper. HST/WFPC2 images of the spiral galaxy NGC 2541, distributed over
13 $V$ and five $I$ epochs within a 47 day window, allowed the
discovery of 34 Cepheid variables and 23 additional variable stars,
none of which previously known.  Photometry for all the stars in the
field was performed independently using a variant of the DoPHOT
program (Schechter et al.  1993, Saha et al. 1994) and the DAOPHOT
II/ALLFRAME package (Stetson 1994); very good agreement (within 0.05
mag) is found between the two sets of photometry.  A subsample of 27
Cepheids with well defined periods between 12 and 47 days, was chosen
to fit $V$ and $I$ PL relations and derive $V$ and $I$ apparent
distance moduli by assuming a Large Magellanic Cloud distance modulus
and mean color excess of $\mu_{LMC} = 18.50 \pm 0.10$ mag and $E(B-V) =
0.10$ mag respectively.  Assuming $R_V = A(V)/E(B-V) = 3.3$, we derive
$A(V)/E(V-I) = 2.45$ based on the extinction laws by Cardelli, Clayton
\& Mathis  (1989), Stanek (1996) and Dean, Warren \& Cousins (1978).
This leads to a true distance modulus to NGC 2541 of $\mu_0 = 30.47
\pm$ 0.11 (random) $\pm$ 0.12 (systematic) mag($ = 12.4 \pm 0.6$
(random) $\pm$ 0.7 (systematic) Mpc, and a total (Galactic plus
internal) mean color excess $E(B-V) = 0.08 \pm 0.05$ (internal error)
mag.

\acknowledgments

We wish to thank Dr Stephane Courteau for providing an IR
Tully$-$Fisher distance to NGC 2541. 
LF acknowledges support by NASA through Hubble Fellowship grant
HF-01081.01-96A   awarded by the Space Telescope Science Institute,
which is operated by the Association of Universities for Research in
Astronomy, Inc., for NASA under contract NAS 5-26555.
The work  presented in this paper  is  based on observations  with the
NASA/ESA Hubble  Space  Telescope,  obtained by  the   Space Telescope
Science Institute, which is operated by AURA, Inc. under NASA contract
No.   5-26555.  Support  for this work   was provided by NASA  through
grant GO-2227-87A from STScI.

\appendix

\section{Other Variable Stars in the NGC 2541 Field}

The 23 stars that did not make the final Cepheid list for the reasons
listed below, are reported in Table A1. Their epoch$-$by$-$epoch DoPHOT
photometry is shown in Tables A2 and A3 (the ALLFRAME photometry is
tabulated for V6, which was not measured by DoPHOT), and their periods
and   phase weighted and intensity averaged magnitudes are listed in
Tables A4 and A5 for DoPHOT and ALLFRAME respectively.  Figures 11 and
12 show $5\sec \times 5\sec$ finding charts and light curves
respectively.

Of the 45 variables identified from the ALLFRAME photometry, all but
one (which was not measured by DoPHOT) are in common with the DoPHOT
variable list.  Five variables selected on the basis of the DoPHOT
photometry did not show convincing evidence of variability from the
ALLFRAME photometry.  For all of them, the ALLFRAME internal errors are
large compared to the magnitude range spanned, and therefore the stars
fall significantly short of the ALLFRAME cutoff criterion for
variability, i.e. $\sigma \ll 1.3$.  Furthermore,  the ALLFRAME
Lafler$-$Kinman plots for these stars are extremely noisy and do not have
a convincing minimum.  Seven more DoPHOT variables were not found in
the ALLFRAME master star list.

For ten of the 44 variables in common between DoPHOT and ALLFRAME that
meet the selection criteria listed in \S 4.1 and \S 4.2, the quality of
the light curves was not deemed good enough for the stars to be
promoted to the `bona fide Cepheids' rank, and used in fitting the PL
relation. The notes to Table A1 list in more detail the reasons for
excluding these variables from the Cepheids sample. These reasons
include the fact that the light curve appears symmetric, the $I$ light
curve is not variable or is not measured, or the amplitude of the $V$
light curve is small, possibly suggesting that the star is contaminated
by a bright companion.

Note that a few of the variables which were not found in the ALLFRAME
photometry do show convincing Cepheid$-$like light curves based on the
DoPHOT photometry (in particular V9). The requirement that only
variables measured as such by {\it both} DoPHOT {\it and} ALLFRAME be
in the final Cepheid list protects us from uncertainties in the
determination of the periods and small ($\le 0.1$ mag) systematic
errors in the photometry (larger errors have already been excluded
based on the comparisons shown in Figure 2).

\section{Secondary Standards in the NGC 2541 Field}

Tables A6a$-$d list positions and DoPHOT photometry  for a set of
bright isolated stars in each chip, used for the DoPHOT and ALLFRAME
comparison shown in Figure 2.

\newpage

\begin{deluxetable}{llll}
\tablecolumns{4}
\tablewidth{0pc}
\scriptsize
\tablecaption{HST Observations of NGC 2541\label{tbl-1}}
\tablehead{
\colhead{Date of Observation} &
\colhead{JD\tablenotemark{a}} &
\colhead{Exposure Time} & 
\colhead{Filter} 
}
\startdata
1994 Dec 28 & 2449714.72809 & 1500 + 1000         & F555W\nl     
            & 2449414.79916 & 1100 + 1400         & F814W\nl 
1995 Oct 30 & 2450020.58781 & 1100 + 1100         & F555W\nl 
            & 2450020.64973 & 1300 + 1300         & F814W\nl 
1995 Nov 05 & 2450027.23003 & 1100 + 1100         & F555W\nl 
            & 2450027.28363 & 1300 + 1300         & F814W\nl 
1995 Nov 13 & 2450035.00566 & 1100 + 1100         & F555W\nl 
1995 Nov 15 & 2450037.01763 & 1100 + 1100         & F555W\nl 
1995 Nov 17 & 2450038.96153 & 1100 + 1100         & F555W\nl 
1995 Nov 20 & 2450041.91476 & 900  + 900  + 260   & F555W\nl 
            & 2450041.97876 & 1100 + 1100 + 260   & F814W\nl 
1995 Nov 22 & 2450044.32303 & 1100 + 1100         & F555W\nl 
1995 Nov 25 & 2450047.40519 & 1100 + 1100         & F555W\nl 
1995 Nov 29 & 2450051.22419 & 1100 + 1100         & F555W\nl 
1995 Dec 03 & 2450055.11055 & 1100 + 1100         & F555W\nl 
1995 Dec 08 & 2450060.33932 & 1100 + 1100         & F555W\nl       
            & 2450060.40402 & 1300 + 1300         & F814W\nl 
1995 Dec 15 & 2450067.10285 & 1100 + 1100         & F555W\nl 
\enddata
\tablenotetext{a}{The Julian Date is given for the middle of the CR$-$split sequence.}
\end{deluxetable}

\clearpage

\begin{deluxetable}{llrllllll}
\tablecolumns{9}
\tablewidth{0pc}
\scriptsize
\tablecaption{Photometric Coefficients\label{tbl-2}}
\tablehead{
\colhead{Chip} &
\colhead{Filter} &
\colhead{ZP$_{DoP}$\tablenotemark{a}} & 
\colhead{AC$_{DoP}$\tablenotemark{b,c,d}} & 
\colhead{ZP$_{ALL}$\tablenotemark{a}} & 
\colhead{AC$_{ALL,1}$\tablenotemark{b,e,d}} &
\colhead{AC$_{ALL,2}$\tablenotemark{b,f,d}} &
\colhead{C2} & 
\colhead{C3} 
}
\startdata
PC  & F555W & 23.878  & -0.9061 $\pm$ 0.0086 (22)  &  24.031 & 0.009 $\pm$ 0.012 (21)  & -0.023 $\pm$ 0.011 (23)  & -0.045 & 0.027\nl
WF2 & F555W & 23.969  & -0.6858 $\pm$ 0.0046 (18)  &  24.042 & 0.007 $\pm$ 0.0085 (24) & -0.012 $\pm$ 0.0090 (23) & -0.045 & 0.027\nl
WF3 & F555W & 23.956  & -0.641  $\pm$ 0.011 (9)    &  24.051 & 0.040 $\pm$ 0.011 (21)  & +0.021 $\pm$ 0.013 (19)   & -0.045 & 0.027\nl
WF4 & F555W & 23.954  & -0.7341 $\pm$ 0.0086 (5)   &  24.027 & 0.023 $\pm$ 0.020 (13)  & -0.009 $\pm$ 0.023 (10)  & -0.045 & 0.027\nl
PC  & F814W & 23.007  & +1.0527 $\pm$ 0.0089 (72)  &  23.137 & 0.013 $\pm$ 0.012 (18)  & +0.032 $\pm$ 0.012 (16)   & -0.067 & 0.025\nl
WF2 & F814W & 23.139  & -0.7536 $\pm$ 0.0094 (85)  &  23.178 & 0.047 $\pm$ 0.0084 (23) & +0.022 $\pm$ 0.0083 (22)  & -0.067 & 0.025\nl
WF3 & F814W & 23.083  & -0.765  $\pm$ 0.014 (23)   &  23.159 & 0.090 $\pm$ 0.010 (22)  & +0.037 $\pm$ 0.010 (20)   & -0.067 & 0.025\nl
WF4 & F814W & 23.084  & -0.786 $\pm$  0.013 (12)   &  23.130 & 0.069 $\pm$ 0.014 (17)  &  0.038 $\pm$ 0.015 (15)  & -0.067 & 0.025\nl
\enddata
\tablenotetext{a}{Errors on both DoPHOT and ALLFRAME zero points are of the order 0.02 mag (Hill et al. 1998).}
\tablenotetext{b}{The number of stars used to determine the aperture correction is shown in parenthesis.}
\tablenotetext{c}{DoPHOT aperture corrections calculated for the 1995 October 30 CR$-$split combined reference epoch (see text for details).}
\tablenotetext{d}{The large difference between DoPHOT and ALLFRAME
aperture corrections is not alarming, and is inherent to the very
different methodologies adopted by the two photometry packages.}
\tablenotetext{e}{ALLFRAME aperture corrections calculated for the first CR$-$split exposure of the 1995 October 30 epoch (see text for details).}
\tablenotetext{f}{As for {\it c} but for the second CR$-$split exposure of the 1995 October 30 epoch.}
\end{deluxetable}

\clearpage

\begin{deluxetable}{llll}
\tablecolumns{4}
\tablewidth{0pc}
\scriptsize
\tablecaption{Comparison of DoPHOT and ALLFRAME for Secondary Standard Stars\label{tbl-3}}
\tablehead{
\colhead{Chip} &
\colhead{\# of Stars} &
\colhead{$\Delta$(F555W)\tablenotemark{a,b}} & 
\colhead{$\Delta$(F814W)\tablenotemark{a,b}} 
}
\startdata
PC  &  24 & +0.015 $\pm$ 0.042  &  -0.027 $\pm$ 0.079  \nl
WF2 &  29 & +0.007 $\pm$ 0.061  &  +0.02 $\pm$ 0.10 \nl
WF3 &  32 & -0.042 $\pm$ 0.042  &  -0.050 $\pm$ 0.050 \nl
WF4 &  29 & 0.013 $\pm$ 0.061  &  -0.023 $\pm$ 0.074 \nl
\enddata
\tablenotetext{a}{The quoted magnitude differences are DoPHOT $-$ ALLFRAME. }
\tablenotetext{b}{The errors represent the rms scatter in the mean.}
\end{deluxetable}

\clearpage

\begin{deluxetable}{llll}
\tablecolumns{4}
\tablewidth{0pc}
\scriptsize
\tablecaption{Comparison of DoPHOT and ALLFRAME for the Cepheids\label{tbl-4}}
\tablehead{
\colhead{Chip} &
\colhead{\# of Stars} &
\colhead{$\Delta$(F555W)\tablenotemark{a,b}} & 
\colhead{$\Delta$(F814W)\tablenotemark{a,b}} 
}
\startdata
PC  &  4 & 0.056 $\pm$ 0.016  &  0.057 $\pm$ 0.025  \nl
WF2 &  17 & 0.010 $\pm$ 0.034  &  0.006 $\pm$ 0.072 \nl
WF3 &  7 & 0.046 $\pm$ 0.063  &  0.026 $\pm$ 0.063 \nl
WF4 &  6 & 0.023 $\pm$ 0.044  &  0.040 $\pm$ 0.037 \nl
\enddata
\tablenotetext{a}{The quoted magnitude differences are DoPHOT $-$ ALLFRAME. }
\tablenotetext{b}{The errors represent the rms scatter in the mean.}
\end{deluxetable}

\clearpage

\begin{deluxetable}{lrrccl}
\tablecolumns{6}
\tablewidth{0pc}
\scriptsize
\tablecaption{Astrometry for the NGC 2541 Cepheids\label{tbl-5}}
\tablehead{
\colhead{ID\tablenotemark{a}} &
\colhead{X\tablenotemark{b}} &
\colhead{Y\tablenotemark{b}} & 
\colhead{RA(J2000)\tablenotemark{c}} & 
\colhead{Dec(J2000)\tablenotemark{c}} &
\colhead{Chip} 
}
\startdata
C1  & 778.40 & 383.03 & 8:14:39.39 & 49:01:30.7 & PC  \nl
C2  & 245.23 & 69.40  & 8:14:43.67 & 49:02:09.5 & WF3 \nl
C3  & 236.86 & 105.11 & 8:14:43.93 & 49:02:07.1 & WF3  \nl
C4  & 634.67 & 648.57 & 8:14:50.68 & 49:02:14.8 & WF3  \nl
C5  & 606.78 & 383.03 & 8:14:39.25 & 49:02:50.3 & WF2  \nl
C6  & 283.15 & 766.84 & 8:14:44.00 & 49:03:07.8 & WF2  \nl
C7  & 300.92 & 152.92 & 8:14:40.82 & 49:02:15.5 & WF2  \nl
C8  & 559.18 & 536.21 & 8:14:39.27 & 49:01:42.8 & PC  \nl
C9  & 526.92 & 602.44 & 8:14:43.80 & 49:00:40.5 & WF4  \nl
C10 & 178.55 & 725.74 & 8:14:44.71 & 49:02:59.9 & WF2  \nl
C11 & 229.14 & 126.18 & 8:14:44.08 & 49:02:05.4 & WF3  \nl
C12 & 764.46 & 560.16 & 8:14:38.75 & 49:03:13.2 & WF2  \nl
C13 & 639.39 & 344.56 & 8:14:38.78 & 49:02:48.5 & WF2  \nl
C14 & 299.21 & 70.01  & 8:14:43.94 & 49:02:14.2 & WF3  \nl
C15 & 640.66 & 491.09 & 8:14:39.49 & 49:03:01.3 & WF2  \nl
C16 & 350.56 & 773.73 & 8:14:43.44 & 49:03:11.7 & WF2  \nl
C17 & 375.52 & 170.34 & 8:14:44.67 & 49:01:25.2 & WF4  \nl
C18 & 353.98 & 340.57 & 8:14:46.58 & 49:02:05.7 & WF3  \nl
C19 & 157.76 & 103.42 & 8:14:43.11 & 49:01:41.7 & WF4  \nl
C20 & 260.79 & 79.30  & 8:14:40.81 & 49:02:07.2 & WF2  \nl
C21 & 622.43 & 605.45 & 8:14:44.62 & 49:00:35.5 & WF4  \nl
C22 & 422.29 & 259.51 & 8:14:44.62 & 49:01:15.2 & WF4  \nl
C23 & 146.71 & 392.39 & 8:14:43.36 & 49:02:28.8 & WF2  \nl
C24 & 480.66 & 445.47 & 8:14:39.82 & 49:01:43.9 & PC  \nl
C25 & 676.48 & 447.00 & 8:14:38.96 & 49:02:59.2 & WF2  \nl
C26 & 436.86 & 484.37 & 8:14:39.76 & 49:01:46.5 & PC  \nl
C27 & 647.39 & 454.90 & 8:14:39.25 & 49:02:58.5 & WF2  \nl
C28 & 277.12 & 687.76 & 8:14:43.66 & 49:03:00.7 & WF2  \nl
C29 & 771.82 & 435.99 & 8:14:38.07 & 49:03:02.8 & WF2  \nl
C30 & 101.78 & 65.83  & 8:14:42.14 & 49:01:58.4 & WF2  \nl
C31 & 146.16 & 536.68 & 8:14:47.27 & 49:01:38.1 & WF3  \nl
C32 & 307.34 & 64.38  & 8:14:44.61 & 49:01:37.7 & WF4  \nl
C33 & 161.59 & 298.56 & 8:14:42.76 & 49:02:21.3 & WF2  \nl
C34 & 315.08 & 416.30 & 8:14:41.99 & 49:02:39.0 & WF2  \nl
\enddata
\tablenotetext{a}{The Variable Stars ID is the same in this and all subsequent tables.}
\tablenotetext{b}{The X and Y coordinate are relative to the 1995 October 30 epoch. For each 800$\times$800 pixels chip, pixel [1,1] is located at the edge of the pyramid.}
\tablenotetext{c}{RA and Declination are calculated using the IRAF task STSDAS.HST$\_$CALIB.WFPC.METRIC, version 1.3.5 (July 1996).}
\end{deluxetable}
\clearpage
{Notes on the individual Cepheids. In what follows the
term `isolated' is used for stars for which no companions are
identified within a three pixel radius. C1: Isolated; period too long
to be determined; phase averaged magnitudes are calculated for P = 65.0
days.  C2:  crowded region; equal brightness companion 2.8 pixels
away.  Period is too long to be determined; phase averaged magnitudes
are calculated for P = 59.0 days. C3:  at the edge of a very crowded
group, even if no other stars are identified within a three pixel
radius.  Period is too long to be determined; phase averaged magnitudes
are calculated for P = 53.0 days. C4: isolated. Period is too long to
be determined; phase averaged magnitudes are calculated for P = 61.0
days.  C5:  two magnitudes fainter companion 2.3 pixels away.  Period
is too long to be determined; phase averaged magnitudes are calculated
for P = 51.5 days.  C6:  two fainter companions 2.8 and 2.0 pixels
away.  Phase averaged magnitudes are calculated for P = 50.7 days.
C7:  2.3 magnitudes fainter companion 3.0 pixels away.  C8:  0.5
magnitudes fainter companion 1.7 pixels away. C9:  at the edge of a
crowded group, even if no other stars are identified within a three
pixel radius.  C10: two companions (one of the same brightness and the other
one magnitude fainter) 3.0 pixels away.  C11:  isolated. C12: two
companions (0.8 and 1.4 mag fainter) within 2 pixels. C13: isolated.
C14: one magnitude fainter companion 1.5 pixels away. C15:  three
companions at least 1 magnitude fainter 2.1 to 2.6 pixels away.  C16:
isolated. C17:  0.6 magnitudes fainter companion 1.6 pixels away. C18:
isolated. C19:  isolated.  C20:  two at least 1 magnitude fainter stars
2.4 and 2.8 pixels away. C21:  isolated. C22: isolated.  C23: two
magnitudes fainter companion 2.8 pixels away.  C24: isolated.  C25: two
at least 1 magnitude fainter companions within 3 pixels.  C26:
isolated.  C27: 1.4 magnitudes fainter companion 2.8 pixels away. C28:
two at least one magnitude fainter companions 2.3 pixels away. C29:
isolated. C30:  isolated. C31:  isolated. C32: isolated.  C33:
isolated. C34: 1.4 magnitudes fainter companion 2.5 pixels away. }

\clearpage

\tablewidth{42pc}


\clearpage

\begin{deluxetable}{lcl}
\tablecolumns{3}
\tablewidth{0pc}
\scriptsize
\tablecaption{Previous Distance Determinations to NGC 2541 and the NGC 2841 Group\label{tbl-11}}
\tablehead{
\colhead{Method} &
\colhead{Distance Modulus} &
\colhead{Reference} 
}
\startdata
{\bf Distance to NGC 2541} & & \nl
Sosie Galaxies  & 29.50 $\pm$ 0.23  & Bottinelli et al. 1985a \nl
B$-$band Tully Fisher Relation  & 29.41 $\pm$ 0.18  & Bottinelli et al. 1984 \nl
B$-$band Tully Fisher Relation  & 29.53 $\pm$ 0.26  & Bottinelli et al. 1985b \nl
B$-$band Tully Fisher Relation  & 29.82 $\pm$ 0.19  & Bottinelli et al. 1986 \nl
IR Tully Fisher Relation & 30.86 $\pm$ 0.30 & Courteau, priv.comm. \nl
Inverse IR Tully Fisher Relation & 30.78 $\pm$ 0.30 & Courteau, priv.comm. \nl
Cepheids & 30.47 $\pm$ 0.23  & This paper \nl
{\bf Other Distance to Group Members} & & \nl
B$-$band Tully Fisher Relation\tablenotemark{a} & 30.54 $\pm$ 0.17 & Bottinelli et al. 1986 \nl
B$-$band Tully Fisher Relation\tablenotemark{b} & 30.44 $\pm$ 0.91 & Bottinelli et al. 1986 \nl
Variety of Secondary Distance Indicators\tablenotemark{c} & 29.53 $\pm$ 0.20 & de Vaucouleurs 1975 \nl
\enddata
\tablenotetext{a}{Distance to NGC 2841}
\tablenotetext{b}{Distance to NGC 2500}
\tablenotetext{c}{Distance to the NGC 2841 group}
\end{deluxetable}%
\clearpage
{Notes on the individual variable stars. V1: isolated,
symmetric light curve, period too long to be determined, phase average
magnitudes are calculated for P=77 days. V2: very crowded region
nearby, small amplitude.  Phase average magnitudes are calculated for
P=60.6 days.  V3: at the edge of a very crowded group, several
companions within five pixels.  Bimodal $V$ light curve. Phase average
magnitudes are calculated for P=46.3 days. V4: fainter companion 3
pixels away. Small amplitude, symmetric light curve; phase average
magnitudes are calculated for P=48.9 days.. V5: one magnitude fainter
companion 2.4 pixels away; not variable from ALLFRAME photometry. V6:
isolated, not found in DoPHOT photometry; magnitudes are derived from
ALLFRAME photometry.  V7: isolated, not found in the $I$ photometry
file. V8:  three fainter companions 2.2 to 2.8 pixels away; not
variable from ALLFRAME photometry. V9: two companions 2.8 pixels away,
$\sim$1 magnitude fainter; not in ALLFRAME photometry. V10:  two
magnitude fainter companion 2.1 pixels away; not variable from ALLFRAME
photometry. V11:  isolated; not in ALLFRAME photometry.  V12:  in a
crowded region, even if no companions are identified within three
pixels; not variable from ALLFRAME photometry. V13: isolated, $V$ light
curve has a flat minimum, possibly due to superposition to a brighter
star. V14: 1.5 magnitudes fainter companion 2.9 pixels away, not in
ALLFRAME photometry. V15: Fainter companion three pixels away.
Symmetric light curve; not sufficient coverage of the light curve in
F814W to derive a reliable $I$ magnitude. V16: Fainter companion 1.5
pixels away. $V$ light curve has flat minimum and small amplitude.
V17: two fainter (0.6 and 1.4 magnitudes respectively) companions 2.3
and 2.9 pixels away; not in ALLFRAME photometry. V18:  flat $I$ light
curve. V19: 0.9 magnitudes fainter companion 2.4 pixels away; not
variable from ALLFRAME photometry. V20: crowded region, three
companions between 3 and 5 pixels away. Not in ALLFRAME photometry,
flat $I$ light curve.  V21: crowded region, three companions between 3
and 5 pixels away.  Eclipsing variable? V22: crowded region, several
companions within 5 pixels. Not in ALLFRAME photometry, $I$ light curve
does not phase with $V$ light curve. V23:  two fainter (.2 and 1.1
magnitudes) companions 2.5 and 2.2 pixels away; not in ALLFRAME
photometry.}

\tablewidth{42pc}


\clearpage


\clearpage

\begin{figure}
\figurenum{2}
\plotone{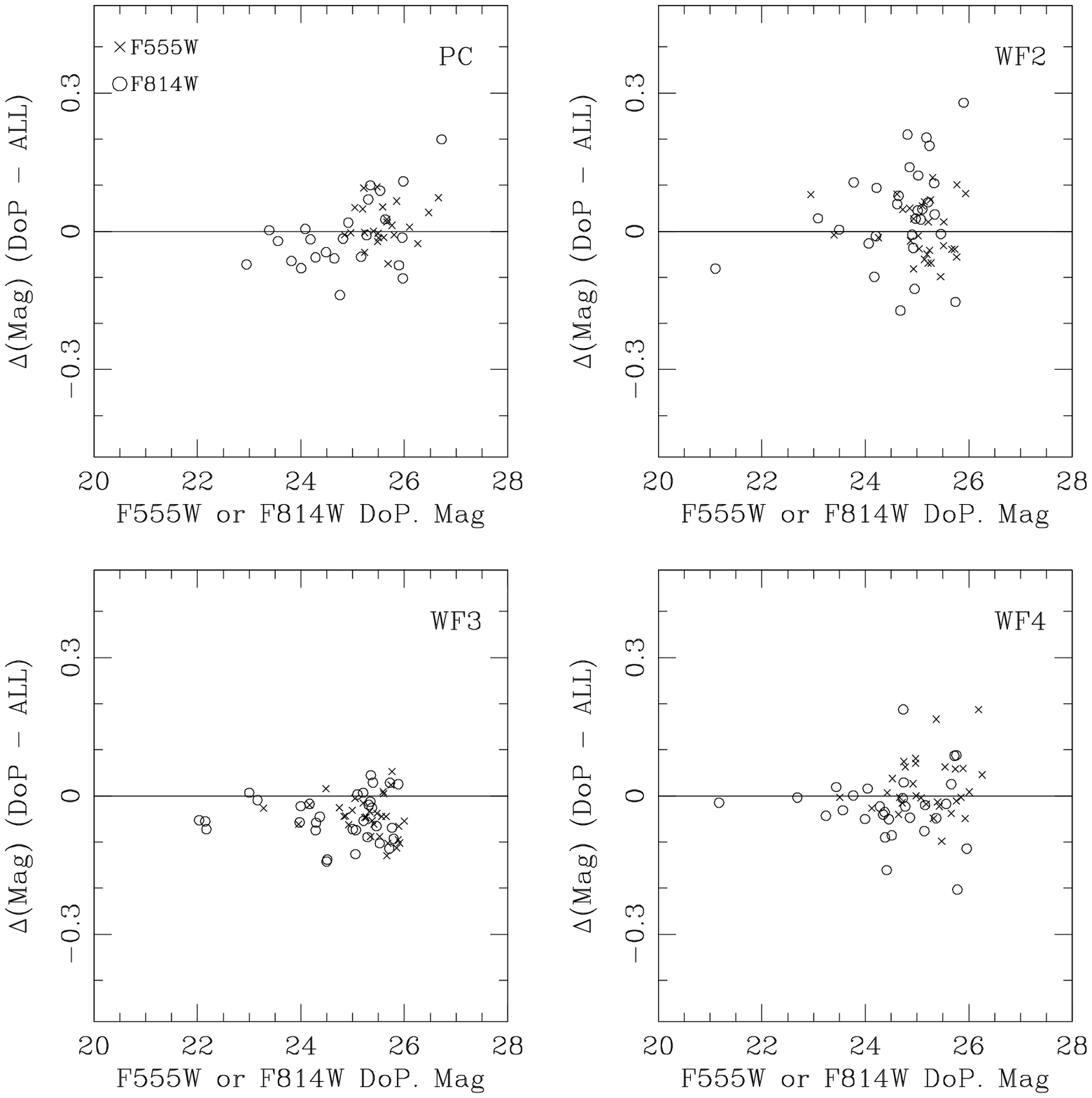}
\caption{Comparison between the DoPHOT and ALLFRAME F555W and F814W
magnitudes  for the secondary standards, whose positions and DoPHOT
photometry are tabulated in Tables A6a-d.}
\end{figure}

\clearpage

\begin{figure}
\figurenum{3}
\plotone{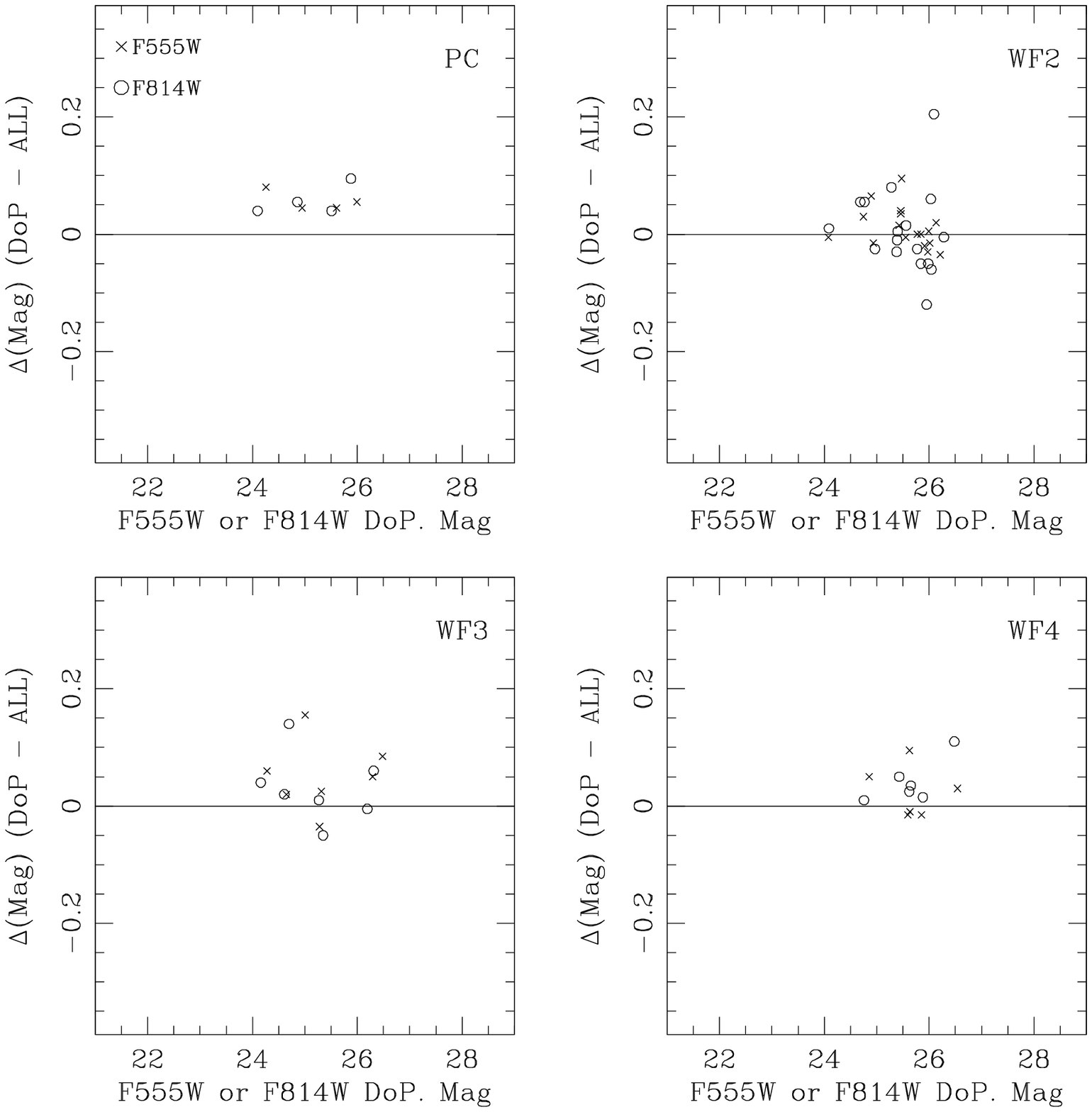}
\caption{Comparison between the DoPHOT and ALLFRAME F555W and F814W
magnitudes  for the Cepheids listed in Table 8. }
\end{figure}

\begin{figure}
\figurenum{6a}
\plotone{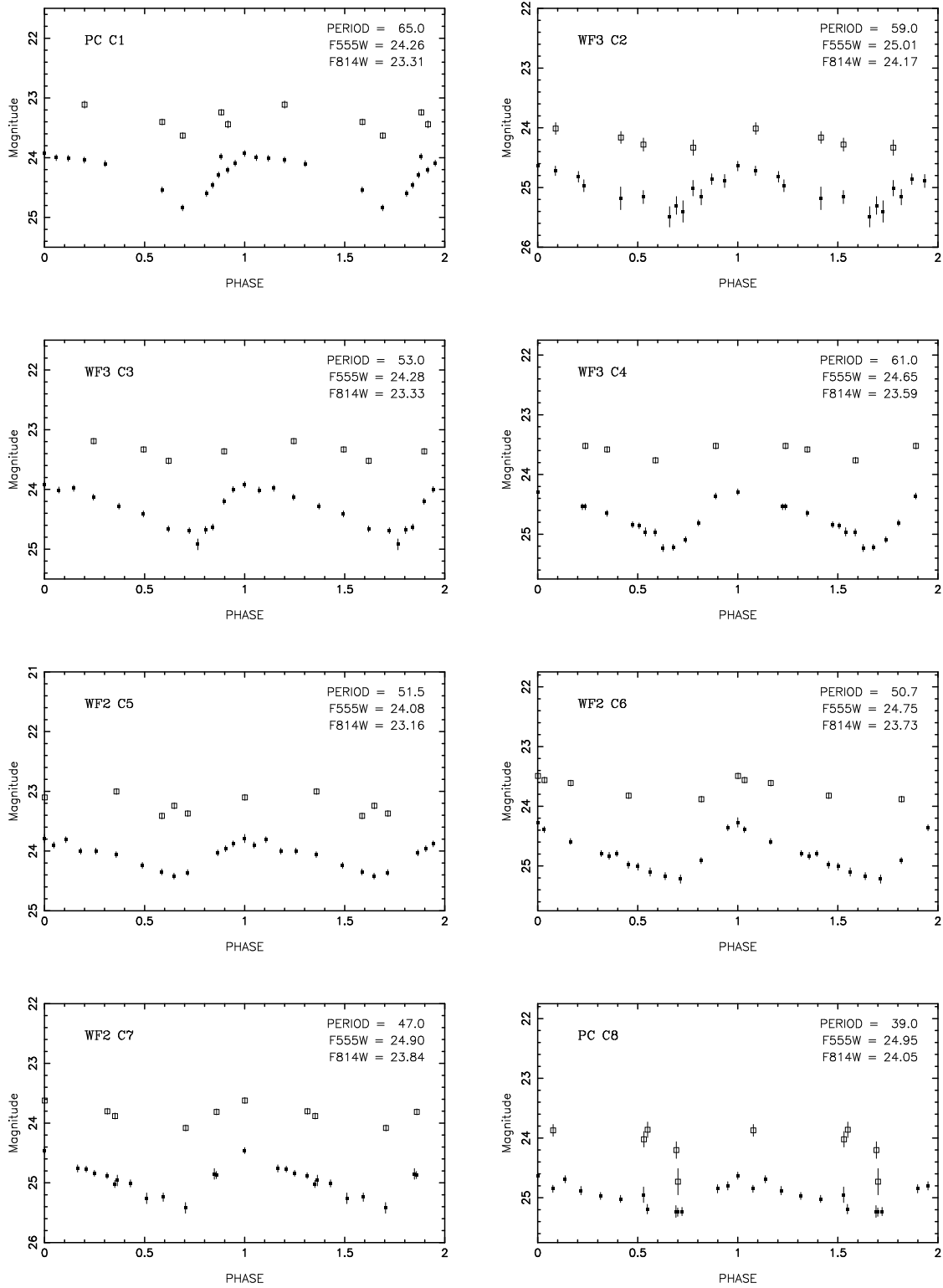}
\caption{DoPHOT light curves for all Cepheids listed in Table 8.
Variables in the plots are labeled with their ID numbers; DoPHOT
periods, F555W and F814W magnitudes (before applying the correction
discussed in \S 4.4) are also reported. Solid and open squares are for
F555W and F814W data points respectively.}
\end{figure}

\begin{figure}
\figurenum{6b}
\plotone{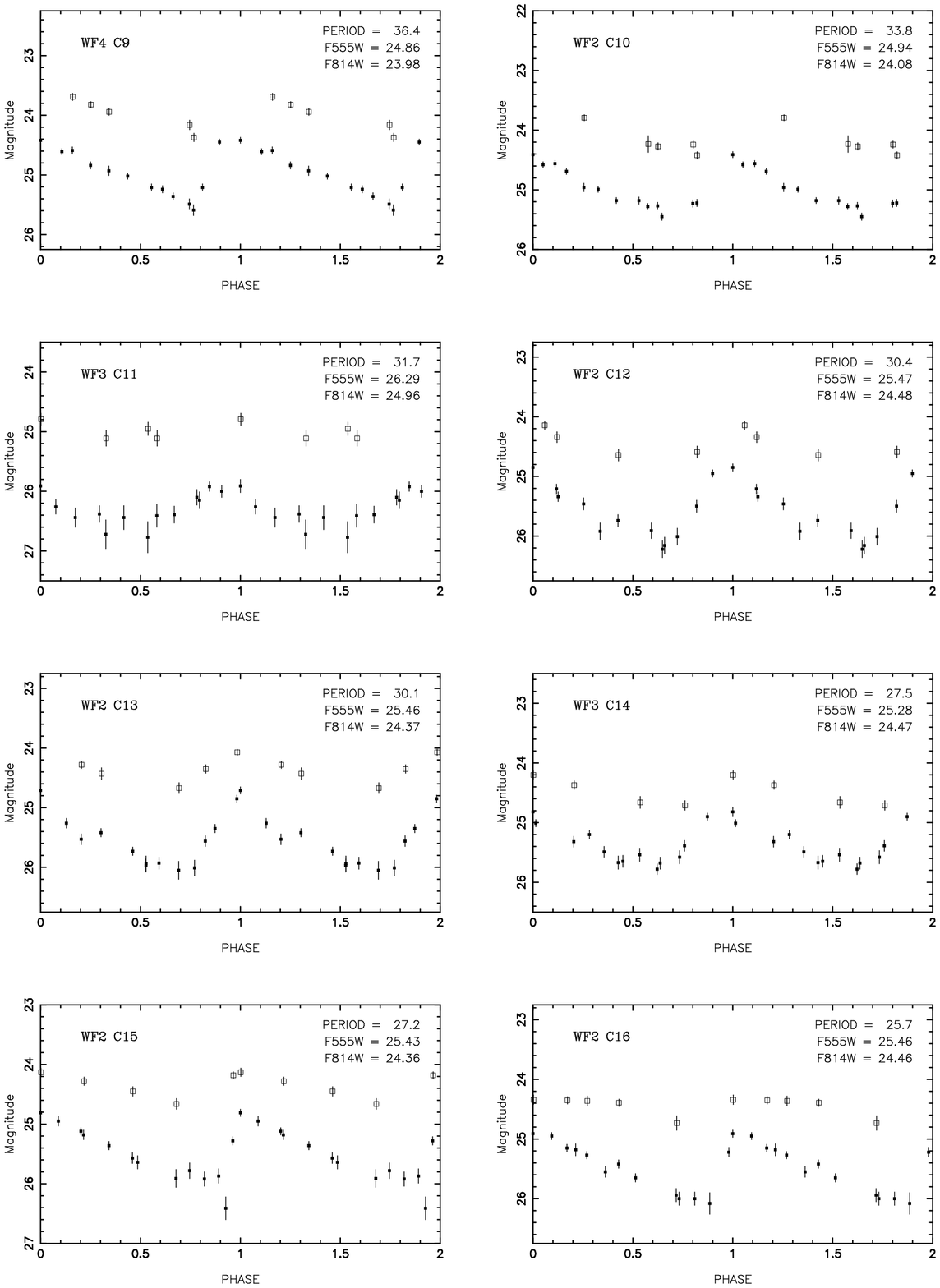}
\caption{As for Figure 6a}
\end{figure}

\clearpage

\begin{figure}
\figurenum{6c}
\plotone{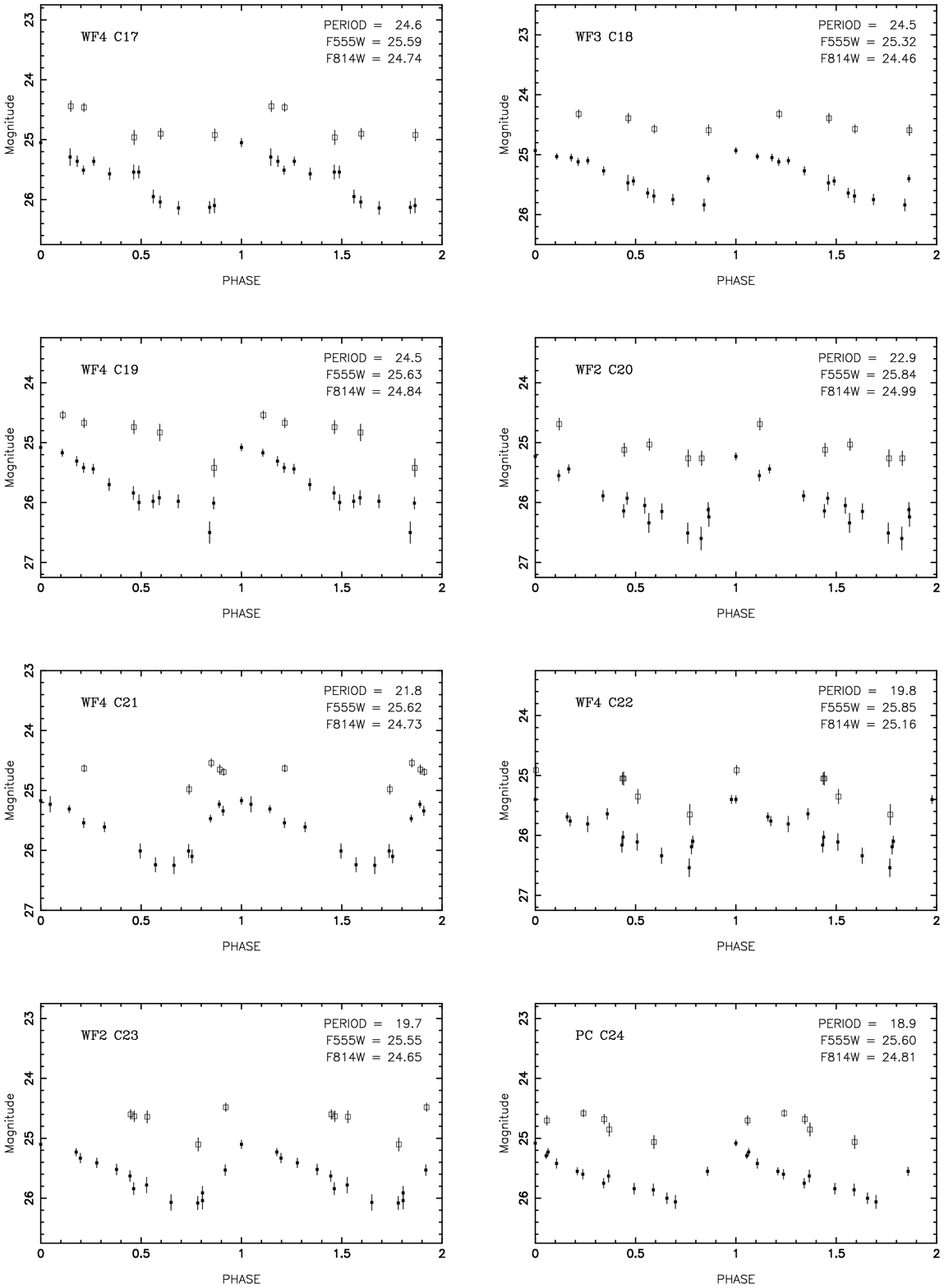}
\caption{As for Figure 6a}
\end{figure}

\begin{figure}
\figurenum{6d}
\plotone{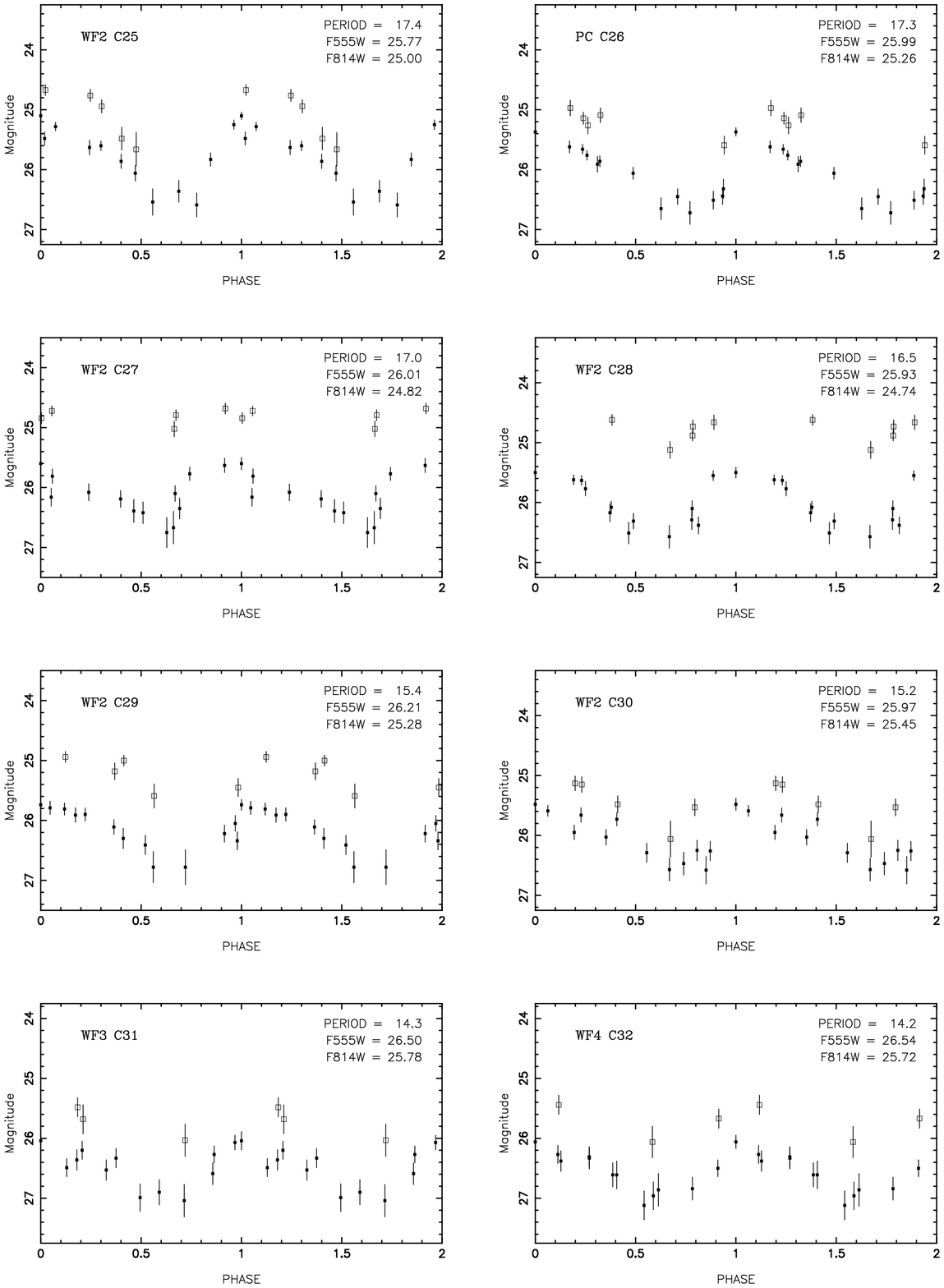}
\caption{As for Figure 6a}
\end{figure}

\clearpage

\begin{figure}
\figurenum{6e}
\plotone{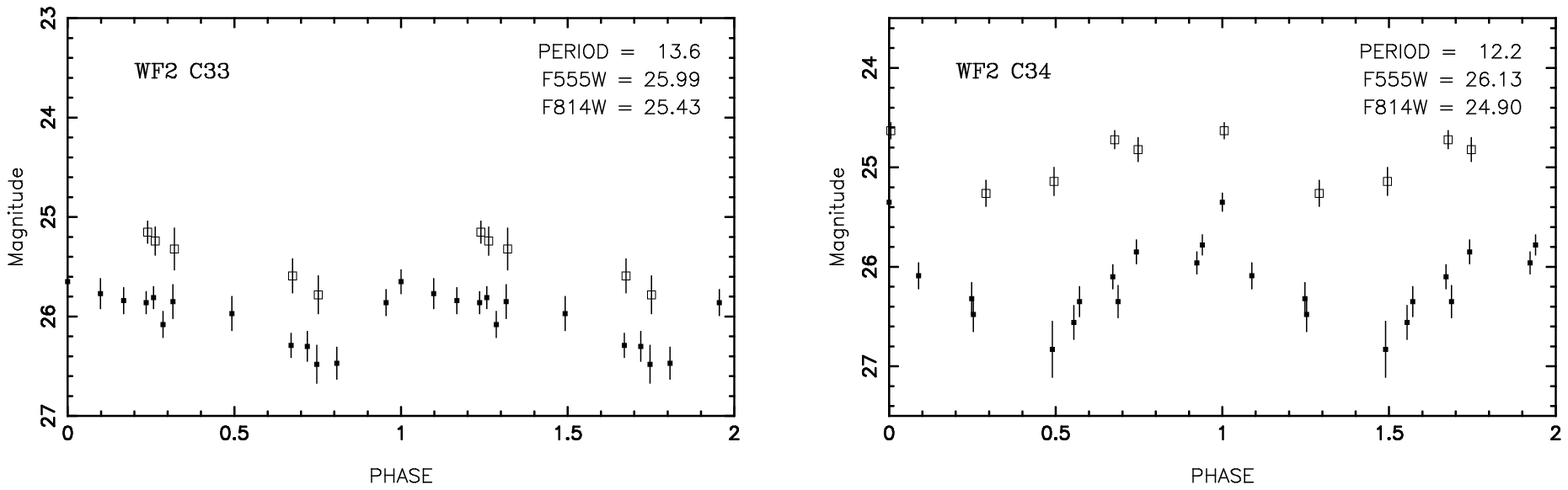}
\caption{As for Figure 6a}
\end{figure}

\begin{figure}
\figurenum{7}
\plotone{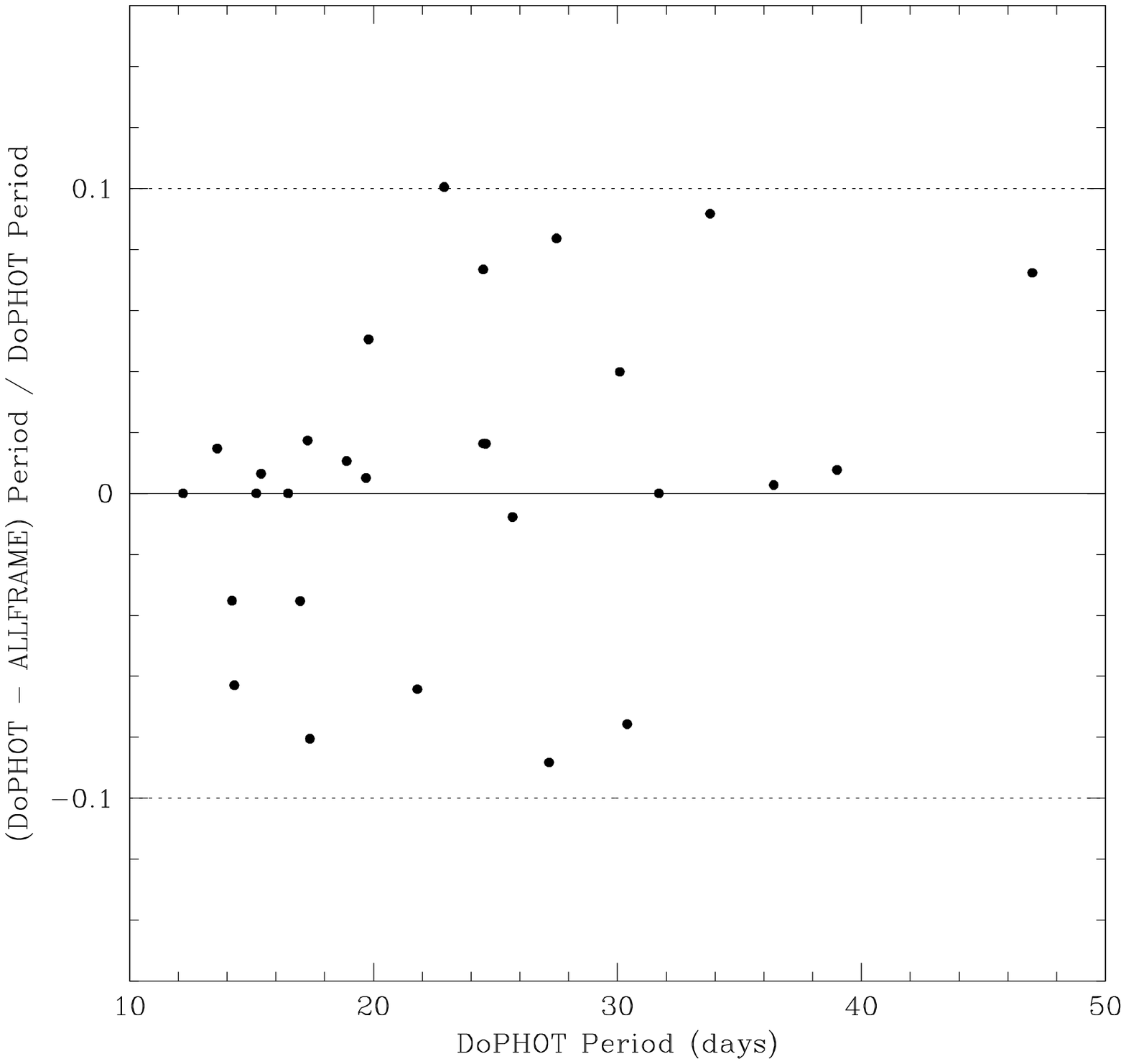}
\caption{Comparison of the DoPHOT and ALLFRAME periods for the 34
Cepheids found in the NGC 2541 field. The dashed lines correspond to a
period difference of $\pm$ 10\%}
\end{figure}

\clearpage




\begin{figure}
\figurenum{10}
\plotone{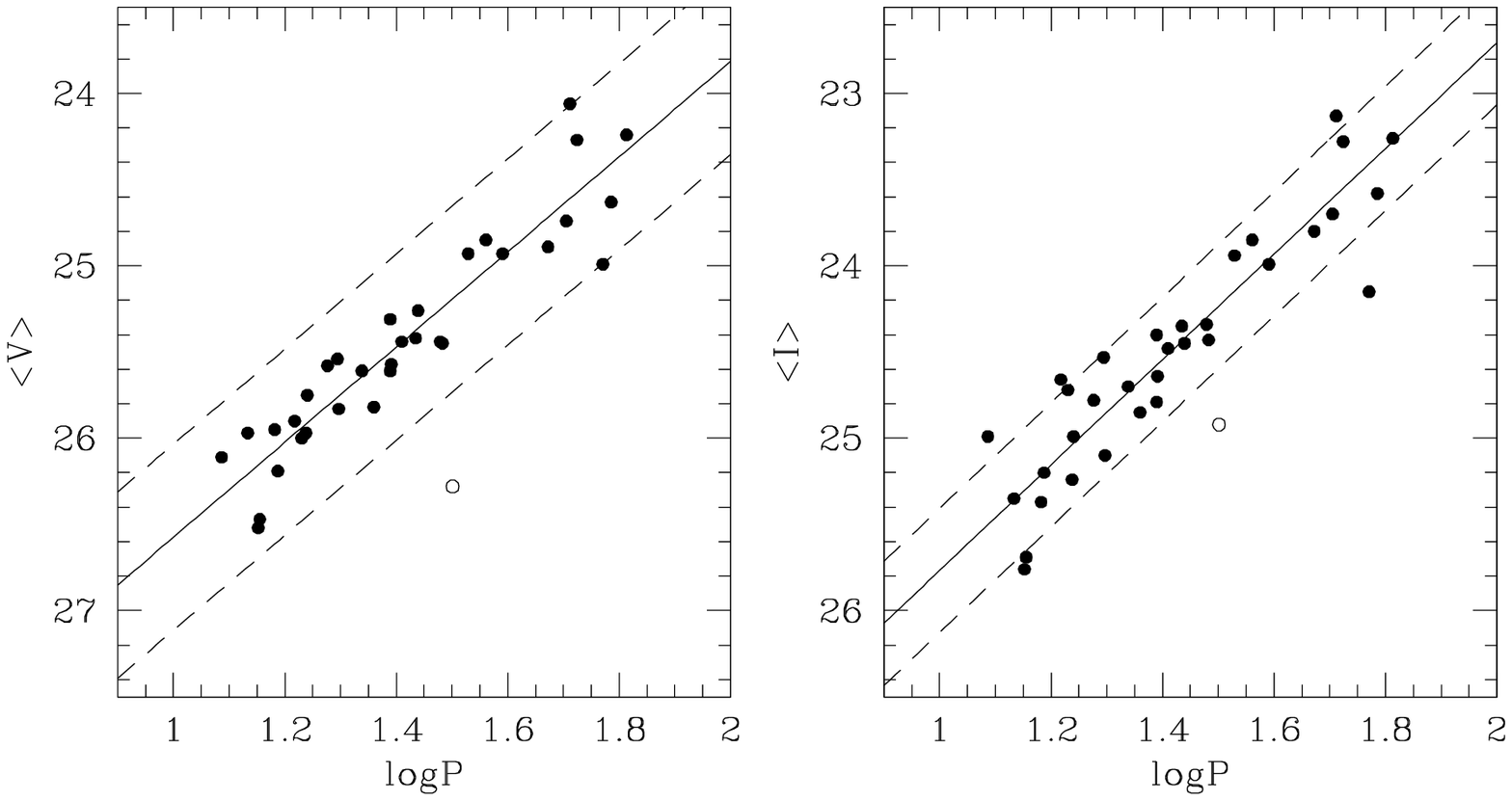}
\caption{$V$ and $I$ PL relations for the sample of 34 Cepheids (see
\S 5 for details).  The best fitting PL relations are shown by the
solid lines. The dashed lines, drawn at $\pm0.54$ mag for the $V$ PL
plot, and $\pm0.36$ mag for the $I$ PL plot, represent the
2$\sigma$ scatter of the best fitting PL relation for the LMC Cepheids 
of Madore \& Freedman (1991).  The only
outlier, C11, is marked by the open circle. Phase weighted DoPHOT
magnitudes are used in the plots.}
\end{figure}

\clearpage


%


\begin{figure}
\figurenum{12a}
\plotone{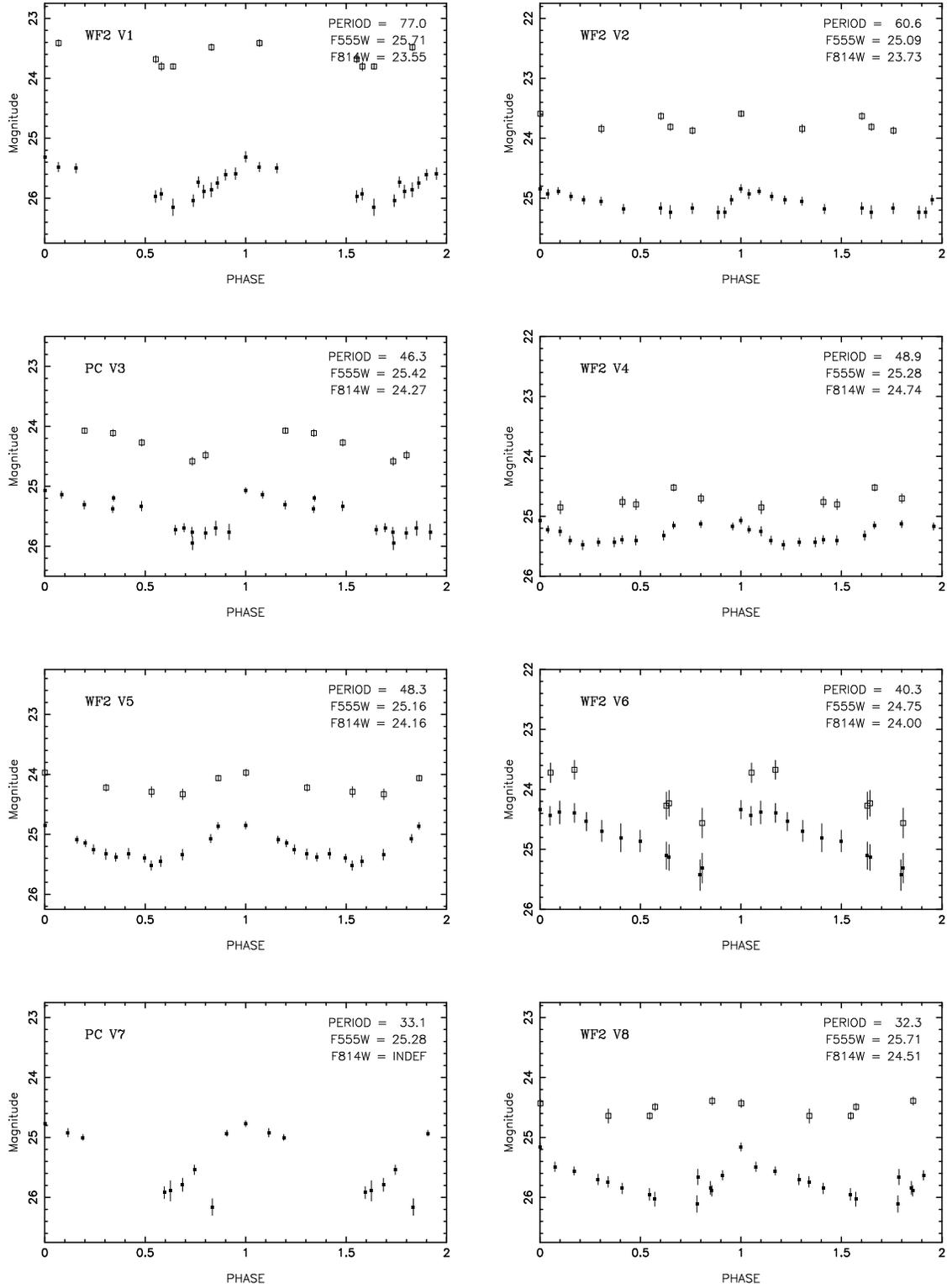}
\caption{Light curves for the variable stars listed in Table A1.
Variables in the plots are labeled with their ID numbers, periods,
F555W and F814W magnitudes (before applying the correction discussed in
\S 4.4). Solid and open squares are for F555W and F814W data points
respectively.}
\end{figure}

\clearpage

\begin{figure}
\figurenum{12b}
\plotone{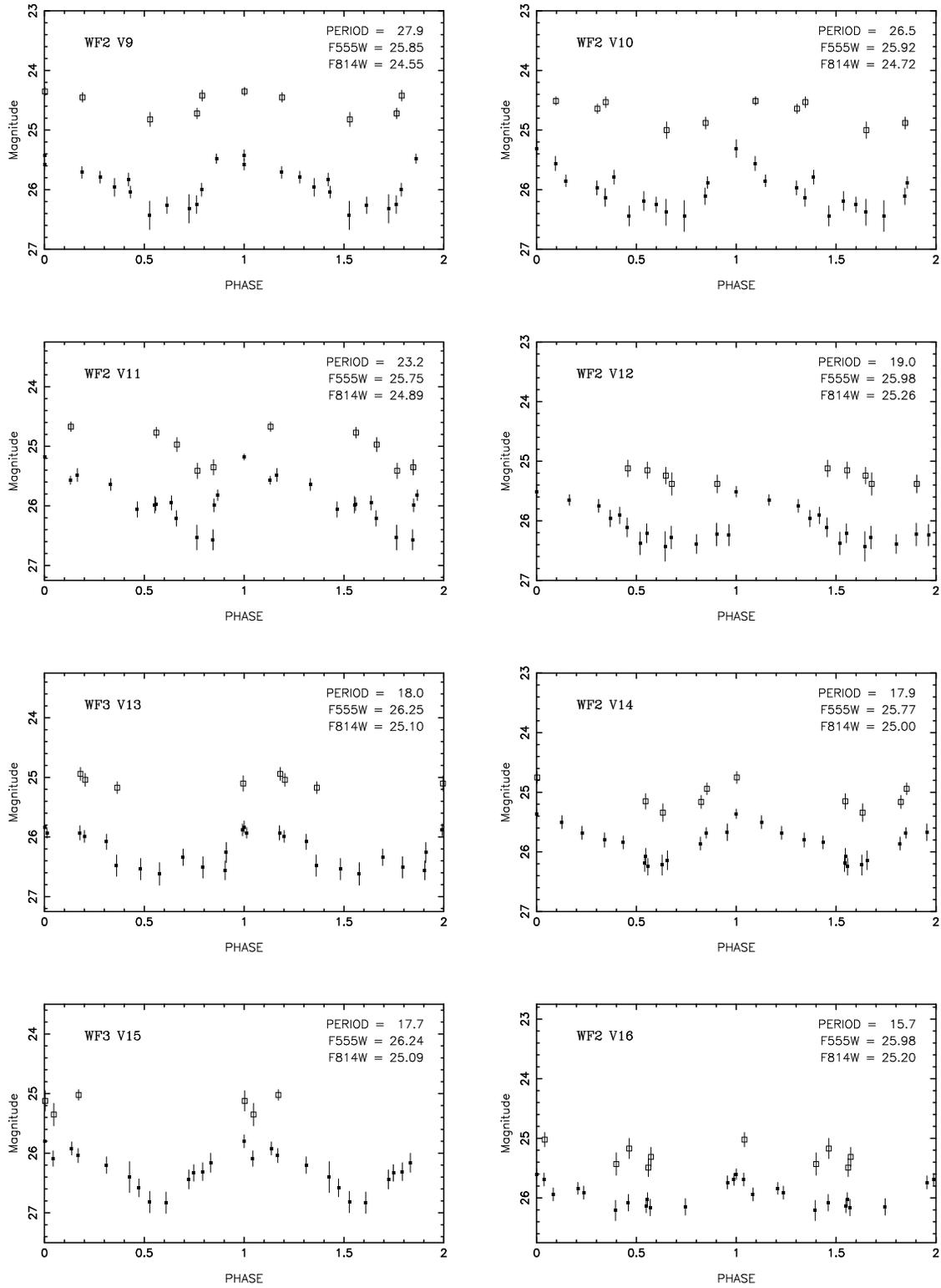}
\caption{As for Figure 12a}
\end{figure}

\begin{figure}
\figurenum{12c}
\plotone{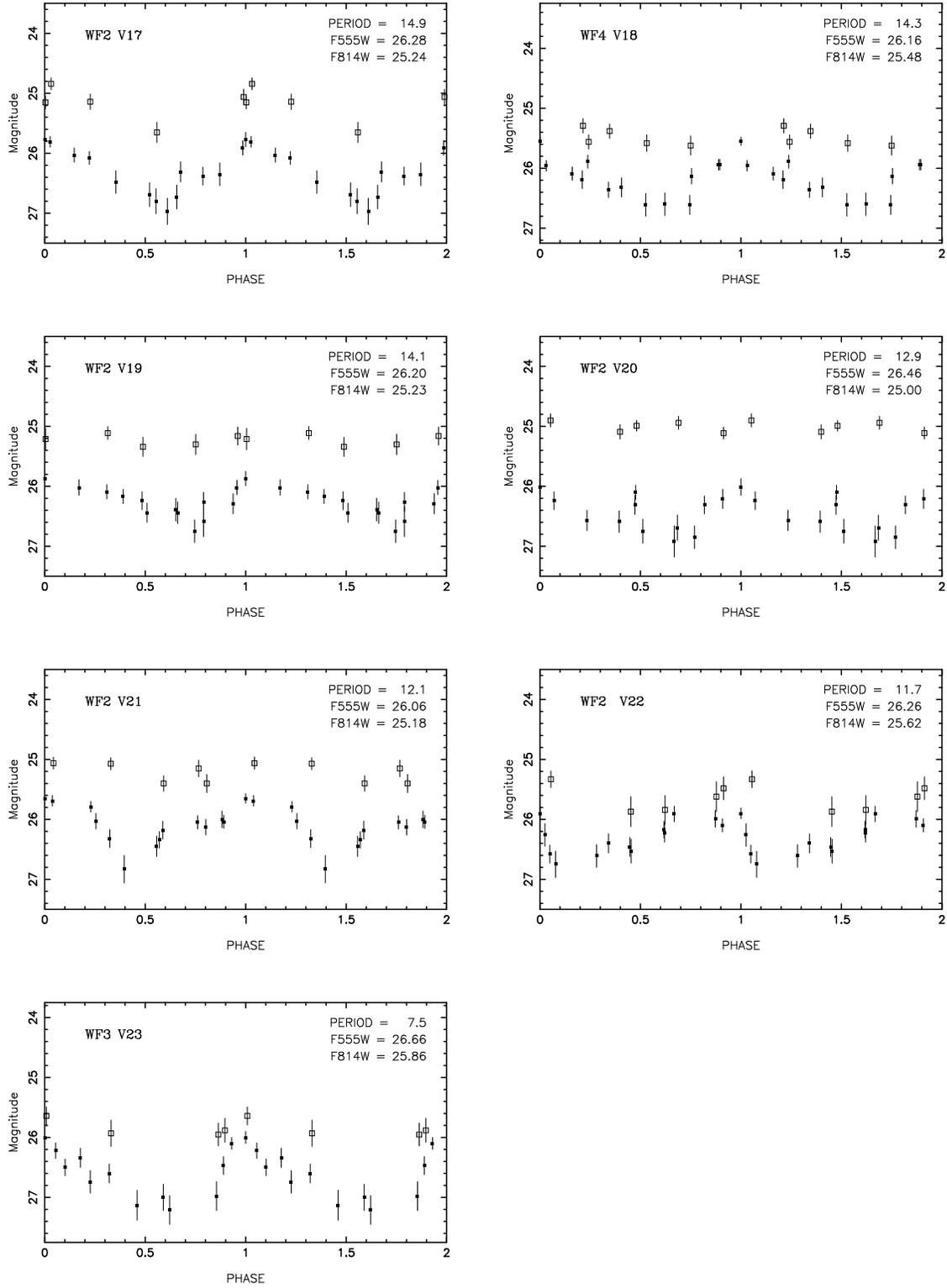}
\caption{As for Figure 12c}
\end{figure}

\end{document}